\documentclass[acmsmall]{acmart}

\AtBeginDocument{%
  \providecommand\BibTeX{{%
    \normalfont B\kern-0.5em{\scshape i\kern-0.25em b}\kern-0.8em\TeX}}}

\setcopyright{rightsretained}
\acmJournal{PACMHCI}
\acmYear{2024} \acmVolume{8} \acmNumber{CSCW1} \acmArticle{42} \acmMonth{4}\acmDOI{10.1145/3637319}

\makeatletter
\gdef\@copyrightpermission{
 \begin{minipage}{0.2\columnwidth}
  \href{https://creativecommons.org/licenses/by/4.0/}{\includegraphics[width=0.90\textwidth]{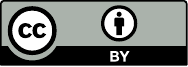}}
 \end{minipage}\hfill
 \begin{minipage}{0.8\columnwidth}
  \href{https://creativecommons.org/licenses/by/4.0/}{This work is licensed under a Creative Commons Attribution International 4.0 License.}
 \end{minipage}
 \vspace{5pt}
}
\makeatother

\usepackage[table]{colortbl}
\usepackage{tikz}
\usetikzlibrary{calc}
\usepackage{zref-savepos}
\usepackage{multirow}
\usepackage[normalem]{ulem}
\usepackage{xcolor}
\usepackage{booktabs,caption}
\usepackage[flushleft]{threeparttable}
\usepackage{setspace}
\usepackage{longtable}
\usepackage{listings}
\usepackage{etoolbox}
\AtBeginEnvironment{algorithmic}{\small}
\usepackage{algorithm}
\usepackage{algpseudocode}
\usepackage[algo2e, ruled,vlined]{algorithm2e}
\SetAlFnt{\small}

\newcounter{indmodel}
\makeatletter
\newenvironment{indmodel}[1][htb]{%
  \let\c@algorithm\c@indmodel
  \renewcommand{\ALG@name}{\footnotesize Individual-level Model}
  \begin{algorithm}[#1]%
    \captionsetup{font=footnotesize}
  }{\end{algorithm}
}
\makeatother 

\newcounter{agmodel}
\makeatletter
\newenvironment{agmodel}[1][htb]{%
  \let\c@algorithm\c@agmodel
  \renewcommand{\ALG@name}{\footnotesize Aggregate-level Model}
  \begin{algorithm}[#1]%
  \captionsetup{font=footnotesize}
  }{\end{algorithm}
}
\makeatother 

\begin{document}

\title{Designing Shared Information Displays for Agents of Varying Strategic Sophistication}

\author{Dongping Zhang}
\email{dzhang@u.northwestern.edu}
\orcid{0000-0001-9825-1411}
\affiliation{%
  \institution{Northwestern University}
  \city{Evanston}
  \state{Illinois}
  \country{USA}}

\author{Jason Hartline}
\email{hartline@northwestern.edu}
\orcid{0000-0001-5505-6819}
\affiliation{%
  \institution{Northwestern University}
  \city{Evanston}
  \state{Illinois}
  \country{USA}}

\author{Jessica Hullman}
\email{jhullman@northwestern.edu}
\orcid{0000-0001-6826-3550}
\affiliation{%
  \institution{Northwestern University}
  \city{Evanston}
  \state{Illinois}
  \country{USA}}

\renewcommand{\shortauthors}{Dongping Zhang, Jason Hartline, \& Jessica Hullman}

\begin{abstract}
  Data-driven predictions are often perceived as inaccurate in hindsight due to behavioral responses. In this study, we explore the role of interface design choices in shaping individuals' decision-making processes in response to predictions presented on a shared information display in a strategic setting. We introduce a novel staged experimental design to investigate the effects of design features, such as visualizations of prediction uncertainty and error, within a repeated congestion game. In this game, participants assume the role of taxi drivers and use a shared information display to decide where to search for their next ride. Our experimental design endows agents with varying level-$k$ depths of thinking, allowing some agents to possess greater sophistication in anticipating the decisions of others using the same information display. Through several extensive experiments, we identify trade-offs between displays that optimize individual decisions and those that best serve the collective social welfare of the system. We find that the influence of display characteristics varies based on an agent's strategic sophistication. We observe that design choices promoting individual-level decision-making can lead to suboptimal system outcomes, as manifested by a lower realization of potential social welfare. However, this decline in social welfare is offset by a reduction in the distribution shift, narrowing the gap between predicted and realized system outcomes, which potentially enhances the perceived reliability and trustworthiness of the information display post hoc. Our findings pave the way for new research questions concerning the design of effective prediction interfaces in strategic environments.
\end{abstract}

\ccsdesc[500]{Human-centered computing~Empirical studies in collaborative and social computing}
\ccsdesc[500]{Human-centered computing~Visualization design and evaluation methods}

\keywords{behavioral game theory; congestion game; strategic decision-making; uncertainty visualization}

\received{January 2023}
\received[revised]{July 2023}
\received[accepted]{November 2023}

\maketitle

\section{Introduction}
Technological affordances enable service providers to leverage historical data and offer users predictions from statistical models to assist decision-making~\cite{gaver1991technology, leonardi2013enterprise}. For example, ad marketplace owners present marketers with predictions on outcomes like clicks or ad placement based on bid amount. Similarly, ride-share and taxi drivers, as well as members of the public attempting to travel from point A to point B, consult predictions on demand surges or traffic congestion from their App or Google Maps, aiding their decisions on where to search for passengers or which route to take. 

These everyday decision scenarios can be viewed as strategic settings of non-cooperative game theory, where multiple \textit{agents} use a shared information display, provided by a \textit{principal}, for decision-making, leading to individual \textit{payoffs} that depend on both the agents' own choices and those of other agents.

In principle, having access to predictions from information displays can benefit agents by providing them with exogenous, payoff-relevant information. However, in practice, the full benefit of this information access may not always be realized. Consider a display predicting a taxi driver's chance of getting a pickup, factoring in the number of other taxis on the road and historical data about where drivers tend to go. Such information displays could guide a driver's decision about where to head next. Yet, taxi drivers attempting to best respond to such a display might be taken aback when the displayed predictions are not realized. The problem arises because decision-making based on a shared information display becomes more challenging in multi-agent strategic settings: the system outcome, formed by combining individual-level decisions, is subject to \textit{distribution shift}~\cite{hardt2016strategic}, where the predicted outcome shown on the display is inaccurate in hindsight because of agents' strategic responses to the displayed predictions.

A principal (i.e., a service provider like a taxi company) faces limited options when confronted with persistent distribution shift resulting from behavioral reactions. Periodically retraining the model is often used in practice. Recent works in machine learning research~\cite{perdomo2020performative, mendler2020stochastic} propose exploring fixed-point solutions within the model retraining space to account for human behavioral responses upon viewing predictions. Drawing inspiration from information design~\cite{bergemann2019information}, a game-theoretical approach involves selectively providing agents with payoff-relevant information to persuade behavioral change. However, the aforementioned solutions have drawbacks, as they can be costly (e.g., model retraining) or rely on stringent theoretical assumptions. We investigate an alternative approach through interface design, examining how design factors of shared information displays influence individual-level decision-making and aggregate-level system outcomes. Additionally, we explore the stability of these dynamics during repeated strategic decision-making over time.

We contribute the design and results of a large online staged experiment using a repeated three-action congestion game based on the search-pickup dynamics of 2.1 million real-world taxi trips. In our experiment, we act as the principal, or the taxi company, while participants assume the role of agents who are taxi drivers. As the principal, the taxi company's objective is to help drivers make good search decisions, ultimately leading to more pickups and improved overall efficiency. To accomplish this objective, the taxi company uses knowledge of how many drivers are on the road and historical data on supply and demand to train a statistical model and present the deduced flows and predicted pickup probabilities of different city districts through a visual interface accessible to all drivers. 

We postulate that two design factors of a shared information display may be particularly influential in shaping agents' decision-making processes. First, we manipulate whether uncertainty in the predicted outcomes is visualized directly, as a salient depiction of uncertainty for predictions may encourage agents to fixate less on a single best response. Second, we manipulate whether \textit{realized prediction error}---the difference between the predicted outcome and what actually happened---is visualized, as understanding the nature of prediction error may aid agents in utilizing an imperfect information display more effectively.  
To study how these factors influence the strategic decision-making of agents, who exhibit realistic variation in their levels of sophistication, we use a staged experimental design, in which we endow each agent with a level-$k$ depth of thinking according to a Poisson Cognitive Hierarchy Model (Poisson-CH model)~\cite{camerer2004cognitive}.
The Poisson-CH model posits that behavioral responses in a strategic setting can be explained by assuming a population of agents who exhibit varying levels of sophistication in how they anticipate other agents' responses. For example, a level-0 agent behaves non-strategically; a level-1 agent attempts to best respond to a population of only level-0s; a level-2 agent attempts to best respond to a fixed mixture over level-0s and level-1s, and so on. 

Our results shed light on the interplay of design elements with strategic outcomes and underscore the challenges of designing shared information displays tailored for agents with varying levels of strategic sophistication. In a pre-registered analysis, we discover that incorporating post hoc decision feedback through visualizing realized prediction error can help more strategically sophisticated agents (i.e., level-2s) make more informed decisions. When decisions are combined to construct the system outcome according to the Poisson distribution used to define the frequency of all levels, design manipulations that improve individual-level decisions can decrease social welfare relative to the system optimal over repeated decisions. Hence, desirable outcomes at the system level can be opposed to those at the individual level. At the same time, the decrease in social welfare is accompanied by a reduction in distribution shift, narrowing the gap between the predicted outcome and the realized system outcome, and hence improving the perceived reliability and trustworthiness of the information display post hoc. We find that these results are robust across two close replications of our experiment, in which we varied the order of decision scenarios and the level distribution of strategic sophistication. We conclude by discussing the implications of our work for future research, focusing on the impact of communicating prediction uncertainty and error on trust and reliance on shared information displays within strategic environments.
\section{Related Work}\label{sec:related}

\subsection{Information Design in Congestion Games}
We study strategic decision-making in congestion games, which represent a broad class of non-cooperative games. Each action of the game represents a congestible good (e.g., local demand or traffic bandwidth) and is associated with a cost function, which incurs a cost that increases with the number or fraction of agents who choose the same action~\cite{roughgarden2005selfish, rosenthal1973class}. In our experiment, the principal's provision and manipulation of displayed information resembles the problem of information design in economics, which studies how a principal can selectively provide payoff-relevant information to influence agents' behavior so as to better achieve the principal's objectives~\cite{bergemann2019information}. Previous work by \citet{das2017reducing} shows that information design can mitigate congestion and improve social welfare in a congestion game. In contrast to selectively releasing information to influence decision-making, our work considers a scenario in which a principal is committed to providing all agents equal access to a shared information display but faces the choice of whether to present agents with prediction uncertainty or post hoc decision feedback on realized prediction error. Our work complements information design in economics by evaluating how the provision of prediction uncertainty and error impacts both individual-level decision-making and aggregate-level system outcomes in repeated decision-making.

\subsection{Strategic Sophistication in Game Theory}
Standard practice in game theory assumes that agents are fully rational and capable of error-free calculations using payoff-relevant information. However, behavioral economists view agents' utility maximization problem through the lens of bounded rationality~\cite{simon1990bounded}, in which agents, typically constrained by limitations in knowledge and computational capacity, tend to satisfice and adapt in their decision-making processes~\cite{jones1999bounded, selten1990bounded, simon1956rational}. 
This perspective has led to the development of several behavioral models, such as Cognitive Hierarchy Models~\cite{camerer2004cognitive}, Prospect Theory~\cite{kahneman2013prospect}, and Quantal Response Equilibrium~\cite{mckelvey1995quantal}, that aim to explain and model the underlying mechanism that dictates behavioral agents' decision-making. Our work adopts a Poisson-based Cognitive Hierarchy Model, which has been extensively studied by empirical game theorists (e.g., \cite{dong2018non, wen2019modelling, fotiadis2021recursive}) by endowing strategic sophistication to agents in a congestion game through a level-$k$ framework. In the Poisson-CH model, the frequency distribution of agents' levels is defined by a Poisson distribution. All agents within the level-$k$ framework are considered to be myopic; they assume they are the most sophisticated agents in action and that all other competing agents are distributed according to a normalized Poisson for levels between 0 and $k-1$. We use the level-$k$ framework to understand how interface design features can affect agents differently depending on their levels of strategic sophistication. 

\subsection{Information Displays for Strategic Decision-making}
Research on uncertainty visualization addresses questions such as how to incentivize uncertainty communication~\cite{hullman2019authors}, how to depict uncertainty information (e.g., \cite{correll2014error, fernandes2018uncertainty, kay2016ish}), and the challenges of evaluating uncertainty displays for decision-making~\cite{hullman2018pursuit}. However, design goals for data and uncertainty visualization have traditionally prioritized individual outcomes, aiming to maximize perception or individual decision quality. Designing interfaces solely to achieve these objectives may not necessarily align with aggregate-level desiderata like greater efficiency or social welfare. Our work addresses novel questions related to how visualizing uncertainty can impact strategic decision-making.

Our work is most related to \citet{kayongo2021visualization}, who proposed the concept of visualization equilibrium. A visualization is in equilibrium if the system outcome observed from agents' decision-making mimics or closely approximates the distribution shown in the display. By studying initial play (i.e., decisions made with no feedback) in a two-action congestion game, they demonstrated how an outcome that a principal might desire for the system, such as a Nash Equilibrium, cannot be achieved by visualizing that outcome but one can estimate an equilibrium by finding a displayed prediction that matches the realized outcome.
Hence, their experiment differs from ours in two important ways: (1) the predictions that the principal provides in their set-up are not constrained by any exogenous information (they can be entirely fictitious), whereas we study a setting in which the principal's predictions are more realistically constrained by real-world taxi behaviors, and (2) they study initial play where agents do not observe any information about the realized outcome (either their choice or the aggregate system outcome) after using a display to make a decision. 

\citet{kayongo2021visualization} also proposed a hypothesis on the impact of visualizing prediction uncertainty on agents' ability to anticipate other agents' actions. They suggested that using a display that can make prediction uncertainty more salient, such as through animated hypothetical outcomes~\cite{hullman2015hypothetical}, may pose challenges for agents in predicting how others will react to the same display. While they provide weak evidence to support this hypothesis, our experiment aims to further investigate its validity through a pre-registered analysis. From a qualitative analysis of participants' reported strategies, they find reports that suggest agents' decision-making behavior might be characterized by varying levels of strategic sophistication when it comes to anticipating other agents' responses, though this analysis is speculative, as they did not measure or endow strategic sophistication in their experiment. In contrast, we explicitly endow different levels of strategic belief according to a Poisson-CH model to understand design factors by level and vary the distribution over levels to check the robustness of our results.
\section{Online Experiments}\label{sec:experiment}
\subsection{Overview}
We conduct a large between-subjects repeated measures experiment on Prolific and two robustness checks in which we replicate the main experiment but change a single assumption. Following a pre-registered\footnote{Pre-registration: \url{https://aspredicted.org/pp8s8.pdf}} analysis plan, we study a three-action congestion game modeled after real-world situations that involve selecting between congestible goods~\cite{roughgarden2005selfish}, in which agents (i.e., participants) act as taxi drivers and are asked to (1) \textit{anticipate} other participants' actions and (2) \textit{decide} where to search for their next ride from three districts using a shared information display. Participants can use the display in this situation to help them maximize their payoffs---or chance of getting a pickup---by accounting for the displayed predictions in their decision-making processes. However, since all participants can access the same information display, whether or not a participant's decision can result in a pickup depends on the decisions of other participants. When more participants choose to search in the same district, the predicted chance of getting a pickup in that district is lower on average. Because agents who use information displays in strategic settings such as congestion games are often long-lived, we observe learning from repeated plays over 15 trials.

The critical component of our congestion game is access to a \textit{counterfactual model} that can compute realistic payoffs for players after decision-making. To inject realism into our setting and to evaluate decisions under exogenous predictions that are not fabricated, we analyze the search-pickup dynamics of 2.1 million Chicago taxi trips and train a counterfactual model that has a functional form of $pickups = f(flow)$. The model is designed to predict 9 AM taxi pickups (i.e., the designated prediction timestamp) in three Chicago Community Areas, each corresponding to an action (i.e., district) in the game, given a discrete flow distribution of drivers going to search over the three districts. We use this model to (1) create predictions modeled after taxi search flows and (2) evaluate participants' decisions.

Decision-making in strategic settings such as the one we study has been found to be well described by assuming that players can vary in how sophisticated they are when anticipating other players' actions~\cite{camerer2004cognitive}. The shared information display we define embodies the assumption that drivers will act \textit{non-strategically}, relying on their prior driving experiences and \textit{without access to the display}. We define these drivers as level-0s (L0s) according to the Poisson-CH model~\cite{camerer2004cognitive}, and simulate the behaviors of these drivers in each decision scenario based on their past search preferences using the taxi data (see Appendix \ref{sec:prior-inference-appendix}).

We endow our study participants with either level-1 (L1) or level-2 (L2) depth of thinking according to a Poisson distribution that characterizes the level frequency over the participant population. Participants who are endowed with L1 belief assume all other participants are L0s, and therefore, the displayed prediction closely resembles the \textit{level-specific outcome} against which they will be scored. L2 participants who believe the population is composed of a fixed mixture of L0s and L1s are expected to make their decisions by combining the information about L0s from the information display with their beliefs about how L1s will attempt to best respond to that display. L2s are scored against a \textit{level-specific outcome} that matches their beliefs, simulated by sampling L0s' decisions from the historical taxi data and L1s' decisions from L1 participants' responses according to L2's endowed belief that there is a fixed mixture of L0s and L1s. Beyond allowing us to study interface effects over realistic variation in participants' levels of strategic sophistication, level endowment enables us to observe how the influence of interface design factors may depend on the extent to which the displayed prediction is aligned with the distribution of behavior that produces the participants' payoffs. 

We employ several specialized terms to delineate our setting and present our findings. For clarity and ease of reference, a glossary of these terms is provided in \autoref{tab:glossary} of Appendix \ref{sec:glossary}.

\subsubsection{Experimental Manipulations}
Similar to \citet{kayongo2021visualization}, we vary whether the prediction uncertainty is visualized directly by showing participants either \textbf{static point estimates} or animated \textbf{hypothetical outcomes plots} (HOPs)~\cite{hullman2015hypothetical}. HOPs are a frequency-based uncertainty visualization technique that presents a finite set of samples from a distribution through a sequence of animated frames. Previous visualization studies suggest that, in some settings, HOPs can yield more accurate probability judgments than error bars~\cite{hofman2020visualizing, kale2018hypothetical, hullman2015hypothetical} or other static methods such as static ensembles and violin plots~\cite{hullman2015hypothetical}. We expect that visualizing static point estimates will lead to less variance in decisions from L1s and L2s, whereas visualizing uncertainty more saliently via HOPs will increase variance in decisions by helping participants recognize that the L0 decisions are not deterministic. 

In each trial, participants are provided with post hoc feedback after decision-making. \textbf{Bandit} feedback only informs the participants whether they received a pickup based on their decision. \textbf{Full} feedback informs the participants if they received a pickup based on their decision and visualizes the \textit{realized prediction error}, or the difference between the predictions they saw prior to making their decisions and the \textit{level-specific outcome} used to evaluate their decisions. Full feedback also visualizes the participant's \textit{anticipation error}, or the difference between the participant's anticipation of the \textit{level-specific outcome} and what is actually realized. By varying the feedback structures, we are primarily interested in assessing whether visualizing realized prediction error can help participants anticipate the prediction error as a result of strategic behavior. 

\begin{table}[ht]
    \caption{Treatment Conditions}
    \label{tab:treatments}
    \begin{tabular}{cccl}
        \toprule
        Interface & Uncertainty Display & Feedback Structure  \\
        \midrule
        1 & Static & Bandit \\
        2 & Static & Full \\
        3 & NetHOPs & Bandit \\
        4 & NetHOPs & Full \\
        \bottomrule
    \end{tabular}
\end{table}

\begin{figure}[ht]
  \centering
  \includegraphics[width=\linewidth]{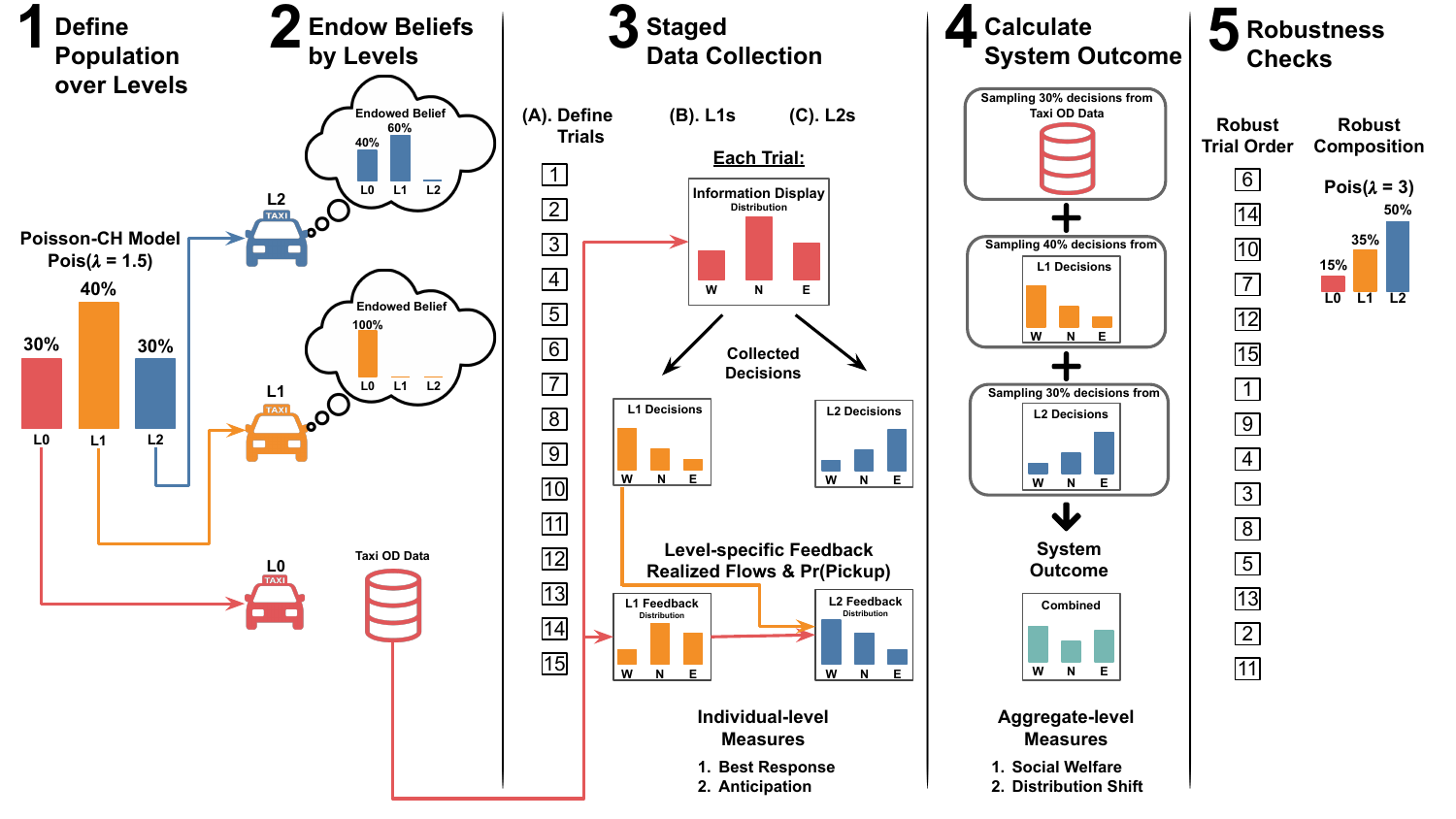}
  \caption{
  Diagram of key features of our experimental design. (1) We define a pseudo-Poisson distribution using a $\text{Poisson}(1.5)$, which includes L0s-L2s. (2) We endow level-specific beliefs by normalizing the pseudo-Poisson distribution for L1s and L2s who are our study participants. (3) We conduct a staged data collection: in each trial, participants use the information display (i.e., shown in \autoref{fig:prediction-display}) to make decisions and then review level-specific feedback (i.e., shown in \autoref{fig:feedback-page}). L1s and L2s complete all trials in the same order, but L1s complete the study before L2s so L1s' responses can be used with that of L0s (i.e., the taxi data) to construct level-specific feedback that aligns with L2s' endowed belief. (4) We calculate the system outcome by combining decisions from L0s (i.e., the taxi data) and L1-L2s (i.e., the collected responses) according to the mixture we used to define the level distribution (i.e., step 1). (5) We conduct two replications of the experiment by varying the trial order and redefining the level mixture of L0s-L2s using a $\text{Poisson}(3)$.
  }
  \label{fig:experiment-diagram}
  \Description{A diagram to demonstrate our experiment design.}
\end{figure}

We vary the uncertainty display and the feedback structure between subjects, resulting in four conditions outlined in \autoref{tab:treatments}. For each condition, we evaluate participants' performance using two individual-level responses: a binary indicator of whether participants \textit{best respond} to the display assuming their level-specific beliefs are correct, and a quantitative measure of their \textit{anticipation error} in anticipating other participants' decisions. Additionally, we simulate a more realistic \textit{system outcome} for each decision scenario. This simulation assumes that the population consists of a fixed mixture of L0s, L1s, and L2s, governed by the Poisson distribution used to define the frequency of all levels. We calculate the \textit{achieved social welfare}, representing the fraction of the total possible social welfare for that decision scenario (i.e., trial) that was achieved, and the \textit{distribution shift}, representing the difference between the displayed prediction and the system outcome. \autoref{fig:experiment-diagram} demonstrates the features of our experimental design. It describes how a distribution over all participants' levels of sophistication gives rise to a data collection arm, where participants' \textit{level-specific feedback} is generated given the endowed level-$k$ beliefs and how we combine decisions to aggregate system outcomes. We describe each step of our experiment with references to the Appendix and supplementary material.

\subsection{Methods}
\subsubsection{Endowing Levels of Strategic Sophistication}\label{sec:level-endownment}
A core aspect of our experimental design is the integration of the Poisson-CH model. In our main experiment, we use a $\text{Poisson}(\lambda)$ where $\lambda = 1.5$ to define the frequency of levels in decision scenarios (i.e., trials) consisting of $N$ participants, based on the findings of \citet{camerer2004cognitive} who analyzed the interplay of level distributions and results of nearly 100 games. Prior work \cite{wen2021modelling, coricelli2009neural, camerer2004cognitive} suggests that human players tend to perform one to two depths of thinking in strategic games. Therefore, we truncate the $\text{Poisson}(1.5)$ and re-standardize its level distribution to a maximum level-$k$ where $k < 3$. Truncating levels in the context of a Poisson-CH model is a common practice within the level-$k$ framework (e.g., \cite{rogers2009heterogeneous, wright2012behavioral, wright2017predicting}). The rationale for excluding levels beyond $k=2$ is based on the characteristics of the Poisson distribution, which determine the population's level frequency. As levels increase beyond a certain threshold, the probability mass associated with higher levels diminishes significantly; thus, as $k$ increases, fewer agents exhibit behaviors associated with levels $k-1$. When $k=4$ and beyond, the behaviors of levels $k-1$ and $k$ become indistinguishable, leading to behavioral convergence. This truncation strategy also aligns with the nature of the Poisson-CH model. It ensures that the model remains practical and interpretable within the given framework.

We demonstrate an example of using a Poisson distribution to define the level composition in \autoref{tab:level-comp-mainExperiment}. Given a $\text{Poisson}(1.5)$ used to define the level distribution of $N$ drivers, we first draw a sample of size $N$ (i.e., counts in row 1 and percentages in row 2) and then normalize the sampled L0-L2s counts to create a 30-40-30 split (i.e., row 3) after rounding the proportions to the nearest tenth to simplify participants' reasoning. This pseudo-Poisson distribution, including L0-L2 drivers, represents the ``true'' population mixture over levels, which is \textit{exclusive} knowledge of the principal used to aggregate the system outcomes by combining decisions using data and the collected responses from the participants. 

Following the level definitions of the Poisson-CH model, our study participants are utility-maximizing but myopic agents whose decision-making is governed by the depth of thinking we endow. As illustrated in \autoref{fig:experiment-diagram}, the non-strategic L0 players in our population are real drivers queried from the taxi data: players who made decisions based on their prior driving experience without using the shared information displays. L1 and L2 players are study participants for whom we endow levels on all study screens that present shared information displays and elicit responses. Specifically, L1 participants are told: 
\begin{quote}
    \textit{All other drivers will NOT consult the display and do NOT know the exact number of competitors in the region. They will drive according to their past driving experiences, which the taxi company has used to create the information display.}
\end{quote}
    
We inform L2s that the other drivers are a mix of L0s and L1s, derived by re-normalizing the pseudo-Poisson distribution used to define all levels, which creates a 40-60 split of L0-L1. Specifically, L2 participants are told:
\begin{quote}
    \textit{[XX]\% ([level-0 count]) of drivers do NOT use the display and do NOT know the exact number of competitors in the region. These uninformed drivers will make decisions according to their past driving experience.}\\
    \textit{[YY]\% ([level-1 count]) of drivers consult the same display as you do, but each of them falsely assumes they are the only person using the display.}
\end{quote}

We evaluate individual-level decisions of participants using \textit{level-specific outcomes} that align with the endowed belief, which we elaborate on in Section \ref{sec:level-specific-feedback}. 

\begin{table}[ht]
    \caption{
    In our main experiment, we use a $\text{Poisson}(1.5)$ to define the level composition for all participants based on the minimum number of drivers across the 15 trial weekdays, which is 598. The $\text{Poisson}(1.5)$ generates a frequency distribution that includes 149 (25\%) L0s, 184 (30\%) L1s, and 139 (23\%) L2s. Notice that this proportion does not sum to 100\% because there are higher levels in the sample, which we omit. We then use the counts of L0-L2s (row 1) to re-normalize the distribution and round to the nearest tenth (row 3). Based on this 30-40-30 split from the pseudo-Poisson, the true level composition for this trial consisting of 598 participants is 180 (30\%), 240 (40\%), and 180 (30\%) L0s, L1s, and L2s (row 4) after rounding again to the nearest tenth to simplify level endowment.
    }
    \label{tab:level-comp-mainExperiment}
    \begin{tabular}{cccc}
        \toprule
        Level-$k$ & 0 & 1 & 2\\
        \midrule
        \#Obs. & 149 & 184 & 139\\
        Percent & 25\% & 30\% & 23\%\\
        Rescaled & \textbf{30\%} & \textbf{40\%} & \textbf{30\%}\\
        True Comp. & 180 & 240 & 180 \\
        \midrule
        Recruit 50\% & 0 & \textbf{120} & \textbf{90}\\
        \bottomrule
    \end{tabular}
\end{table}
\subsubsection{Tasks and Rewards} 
We create the decision scenarios that constitute the trials of our experiment by sampling 15 unique but homogeneous weekdays from the taxi data used to train the counterfactual model. Participants in the experiment are repeatedly presented with the question: ``Where should I find my next pickup at 9 AM?" based on information displays that render both the deduced search flows and the predicted pickup probabilities, assuming all drivers act according to their past driving experiences. Although our decision scenarios have the same time setting, the search-pickup dynamics presented in the information displays vary because each reflects decisions of a unique set of $N$ drivers on that specific historical weekday (see Appendix \ref{sec:reduce-heterogeneity-appendix}). We keep the time of the decision scenarios fixed at 9 AM on weekdays so that participants in the experiment are in a position to learn from repeated decisions, similar to how a taxi or ride-share driver might form a mental model of what dynamics to expect during a given driving time frame.

\begin{figure}[ht]
  \centering
  \includegraphics[width=.8\linewidth]{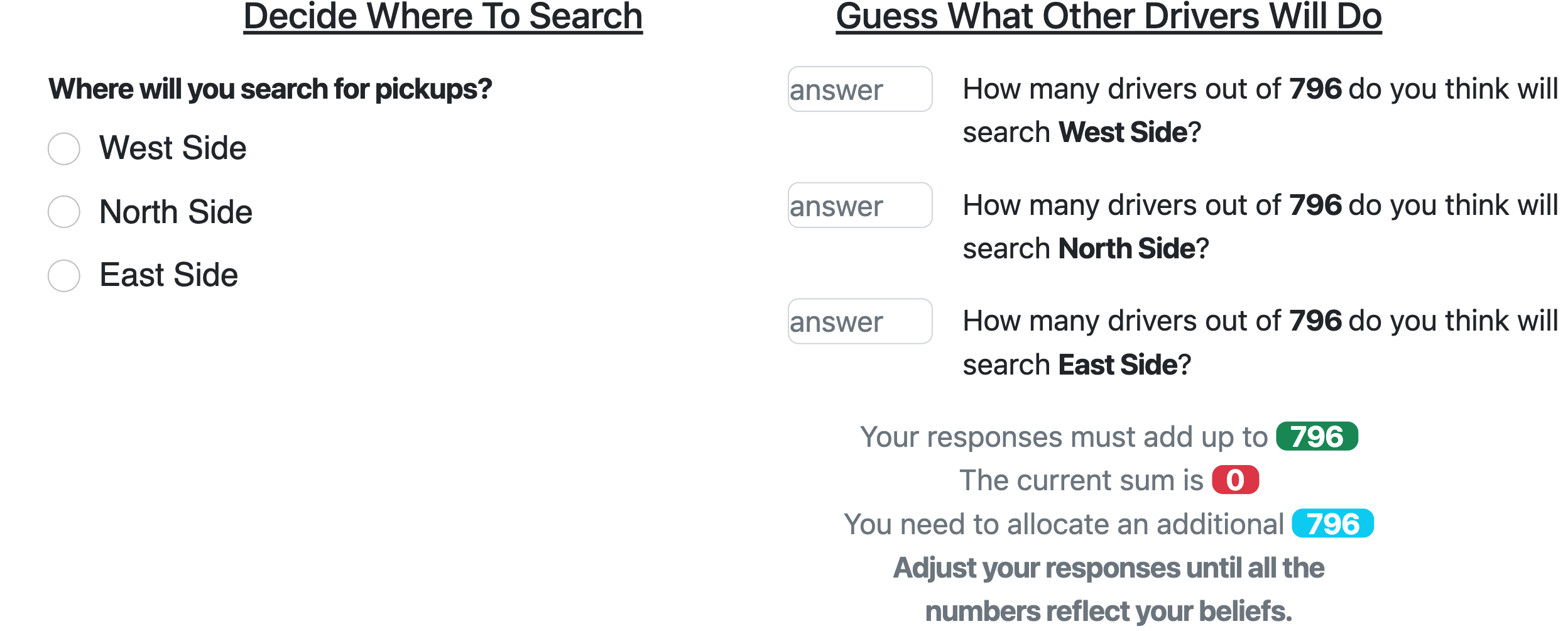}
  \caption{
  The interface used to collect participants' decisions. When providing anticipation, after a participant provides guesses for two districts, the interface imputes the flow of the last district to ensure proper summation of the total number of drivers of the decision scenario. The interface dynamically updates the current flow sum and the amount of flow to be allocated or removed if the elicited flows do not sum to the correct total.
  } 
  \label{fig:elicitation-interface}
  \Description{A demonstration of the elicitation interface.}
\end{figure}

In each trial, we elicit participants' decisions and their corresponding anticipation of other participants' actions. Participants are asked to (1) \textit{decide} where to search by selecting a district and (2) \textit{anticipate} what other drivers will do by entering the number of drivers that they think will search in each district according to the endowed level. As shown by the elicitation interface in \autoref{fig:elicitation-interface}, participants select from multiple choice options and use a dynamic input form based on previous work on eliciting Dirichlet distributions~\cite{chaloner1987some,o2006uncertain} to provide anticipated flows. Each participant receives a base pay of $\$2$ and a bonus of $\$0.2$ for each trial in which their selected strategy resulted in a pickup.

\subsubsection{Generating Shared Information Displays}\label{sec:generate-prediction-display}
For each trial, the information display presents flows going into the three districts by deducing decisions using the search preferences of real taxi drivers (i.e., L0s) involved in the decision scenario. These deduced flows are then used to predict pickup probabilities with the counterfactual model. As decisions based on real drivers' search preferences are subject to uncertainty, we use simulations to create a distribution of hypothetical outcomes. Each outcome is used as a frame for the animated display, and these frames are aggregated to produce the static display.

\paragraph{\textbf{Simulated Flows and Predicted Payoffs}} \label{sec:simulated-flow} 
We first identify a set of $N$ candidate drivers involved in a decision scenario who would be able to search our districts at 9 AM from their current location (i.e., see \textit{trace dyad} in Appendix \ref{sec:prior-inference-appendix}). We then consult each candidate driver's conditional search prior, which is encapsulated in the driver's \textit{search dyad} $V\xrightarrow{N}S$, where $V$ is the drop-off district of the previous trip, $S$ is the pickup district of the consecutive trip, and the weight $N$ is the number of occurrences of the pickup pattern (i.e., see \textit{search dyad} in Appendix \ref{sec:prior-inference-appendix}). Because a \textit{search dyad} summarizes a driver's search preferences from the current location based on her pickup history over the past ten days, we deduce a driver's search decision by sampling $s$ using $n$ as weight. We combine sampled decisions to deduce the flow of each district and address uncertainty by replicating this procedure 1,000 times, so each district has a distribution of simulated hypothetical outcomes, including the deduced flow and the resulting pickup probabilities predicted by the counterfactual model. We provide detailed descriptions of our counterfactual model in Appendix \ref{sec:counterfactual-appendix}.

\paragraph{\textbf{Visualizing Predictions as Networks}} 
The action set (i.e., possible choices) of our congestion game contains three districts forming a traffic network describing flows. We present these predictions in the form of node-link diagrams of an egocentric network, which is a common representation of flow data. Our design choice is governed by the fact that the information display communicates two important forms of payoff-relevant information to participants: (1) deduced flows and (2) predicted pickup probabilities. Nodes representing districts are labeled with district names\footnote{Each district within the action set corresponds to one of three Chicago Community Areas: North Loop as the ``North District," West Loop as the ``West District," and the Loop as the ``East District." For a more detailed description of the action set, see Appendix \ref{sec:action-sets-appendix}.} with the corresponding predicted pickup probability as text above the node. This value is also encoded as node size and hue. Links connecting the ego and nodes of districts (i.e., alters) represent deduced flows, where the amount of flow is encoded by edge width. We position the nodes on the display to resemble an outgoing star~\cite{lusher2013exponential} using a force-directed layout algorithm~\cite{eades1984heuristic}. 

\begin{figure}[ht]
  \centering
  \includegraphics[width=\linewidth, keepaspectratio]{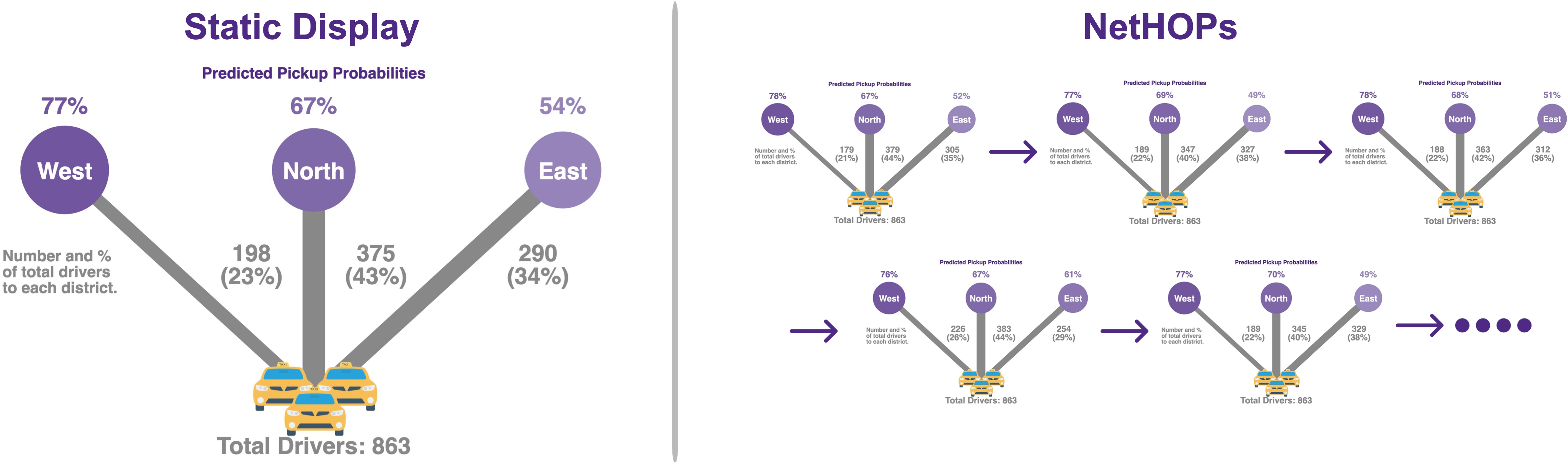}
  \caption{
  Examples of the information displays used for decision scenarios varied by uncertainty quantification. \textbf{Right}: NetHOPs, which render 1,000 hypothetical outcomes, are presented in a looping animation with an animation speed of 0.2 seconds per frame using a fixed force-directed layout (i.e., anchoring $\alpha = 1$). This approach follows suggestions from \citet{zhang2021visualizing} that best support node-attribute and link-attribute tasks. \textbf{Left}: Point estimates where the rendered payoffs are the weighted averages of all simulations.
  }
  \label{fig:prediction-display}
  \Description{Example information display.}
\end{figure}

We randomly vary the display types between participants. As shown in \autoref{fig:prediction-display}, some participants are assigned Network Hypothetical Outcome Plots (NetHOPs)~\cite{zhang2021visualizing}, which depict prediction uncertainty more saliently by showing simulated network realizations. Following suggestions from \citet{zhang2021visualizing} on how to tune NetHOPs' visualization parameters to support node-attribute and link-attribute tasks effectively, we render 1,000 hypothetical outcomes in a looping animation with an animation speed of 0.2 seconds per frame using a fixed layout (i.e., anchoring $\alpha = 1$). The remaining participants are assigned a static node-link diagram of point estimates, in which the deduced flows and predicted pickup probabilities visualized are the weighted averages over the 1,000 hypothetical outcomes.

\subsubsection{Evaluating Decisions and Generating Level-Specific Outcome}\label{sec:evaluate-decisions}
\paragraph{\textbf{Staged Experiment Design}} 
Our desire to incorporate both realism and control into the strategic setting, especially by introducing variation in players' levels of strategic sophistication, presents a challenge: to evaluate decisions by higher-level participants (i.e., L2s), we need access to decisions from the lower levels. While this is trivial for L0s as their decisions are drawn from the taxi data as shown in \autoref{fig:experiment-diagram}, scoring and giving feedback to L2 decisions requires access to L1 decisions. Consequently, we designed a staged data collection procedure in which L1 participants first complete the series of trials and are scored according to L1's endowed beliefs, then L2s complete the same series of trials but are scored according to L2's endowed beliefs.

There are several implications of this experimental design choice. Because the post hoc decision feedback participants receive is \textit{level-specific}, we can study individual-level performance under the assumption that players' beliefs are fixed and contrast the impact of interface manipulations on players of varying levels of sophistication. By combining L0, L1, and L2 according to the Poisson distribution used to define the frequency of all levels, we can also calculate aggregate-level system outcomes under the assumption that players are myopic (i.e., unaware of other players at the same level as them and above). Note that this requires a fixed trial order where all agents experience the scenarios in the same order so that the aggregate results are not confounded by differences stemming from prior decision-making. The primary limitation of the staged design is that we cannot know to what extent receiving decision feedback based on the ``true" population mixture, including players of all levels, would change the patterns of behavior that we observe.

\paragraph{\textbf{Computing Level-specific Feedback}}\label{sec:level-specific-feedback}
We generate \textit{level-specific outcomes} that align with participants' endowed levels to score their decisions and provide feedback in each decision scenario. Recall that an L1 player believes herself to be the most sophisticated player and assumes that her competitors are all L0s. Therefore, L1s' decisions are evaluated against a \textit{level-specific outcome} created by combining the decisions of L0s, which is the distribution of search flow and the corresponding pickup probabilities we used to fit the counterfactual model. Similarly, since an L2 player believes that she is playing against a combination of L0s and L1s, we generate L2s' \textit{level-specific outcome} by sampling the decisions of L0s (i.e., from the taxi data) and L1s according to the proportion of L0s and L1s endowed to L2s (see Section \ref{sec:level-endownment}), which is possible as a result of the staged data collection. To reduce sampling error, we repeat the sampling procedure 1,000 times and summarize the samples by computing the expected flow distribution. We use this distribution to predict each district's pickup probabilities by the counterfactual model and to evaluate L2's decisions.

\begin{figure}[ht]
  \centering
  \includegraphics[width=\linewidth]{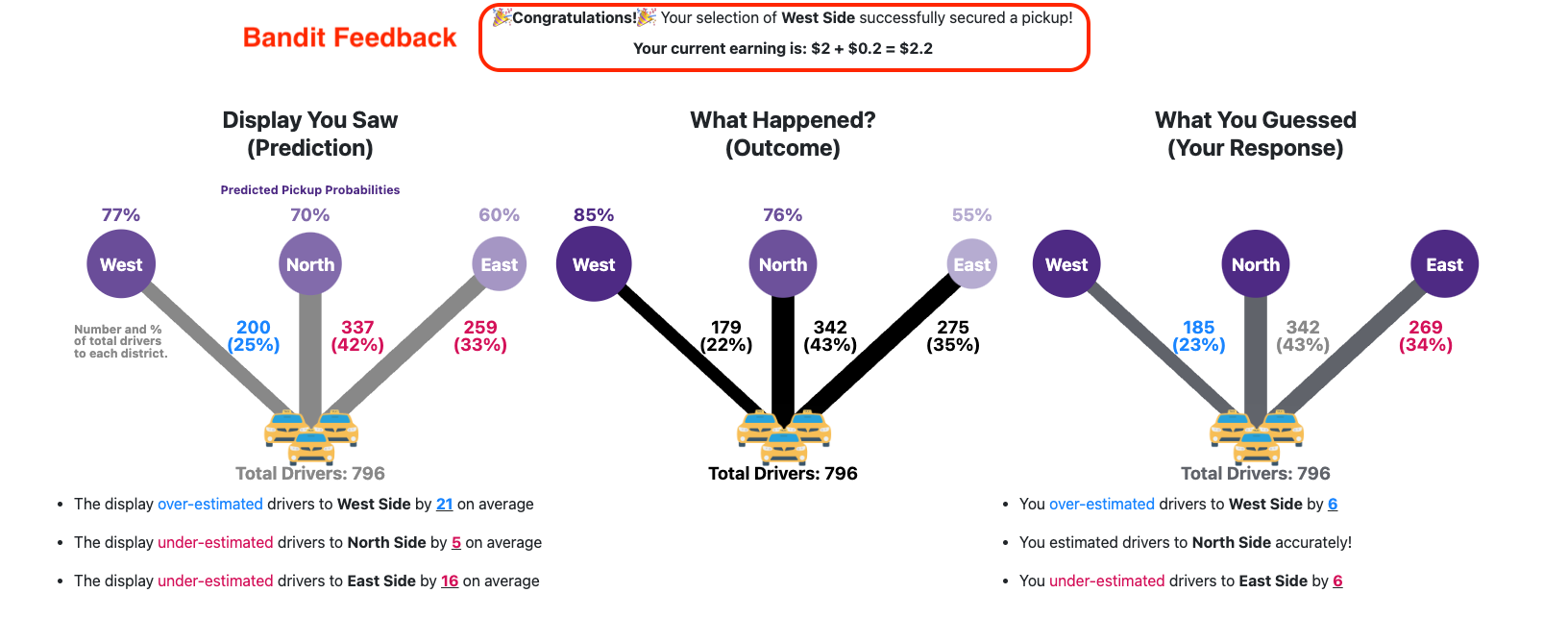}
  \caption{
  An example of bandit and full feedback provided after participants submit their responses. Both feedback types display the decision result and current study compensation (i.e., highlighted in red). Bandit feedback exclusively provides this information. Full feedback additionally presents three visualizations, reminding participants of the prediction they used (left), the \textit{level-specific outcome} (middle), and their anticipation submitted with the decision (right).
  }
  \label{fig:feedback-page}
  \Description{A demonstration of full decision feedback provided to participants.}
\end{figure}

\paragraph{\textbf{Feedback Structure}}
We evaluate two types of decision feedback: \textit{bandit} and \textit{full}. In the full feedback conditions, participants are presented with the \textit{level-specific outcome} along with a reproduction of the predictions and their anticipated flows, as illustrated in \autoref{fig:feedback-page}. To enhance the visibility of realized prediction and anticipation error, we use blue color to denote over-estimation and red color to denote under-estimation in the labels of the feedback displays and the accompanying summarizing text. In the bandit feedback condition, participants receive a subset of the feedback information provided in the full feedback condition, focusing solely on the decision results, as highlighted in \autoref{fig:feedback-page}.

\subsubsection{Experiment Procedure}
Participants are directed to our study interface\footnote{Interface screenshots: \url{https://github.com/dpzhang/shared-info-display}} by Prolific. They first see a welcome page with the estimated study completion time and compensation. Participants are instructed to complete the study in a single session using Google Chrome on a large-screen device. If participants agree to the terms and click a confirmation button, they are randomly assigned to one of the four treatment conditions shown in \autoref{tab:treatments}. 

Upon enrollment, participants must review four detailed instruction pages. These instructions provide a comprehensive overview of the study, encompassing the experimental setting and specific guidance on using the assigned display and feedback for decision-making. After reviewing the second instruction page, participants encounter a multiple-choice question designed to assess their understanding of the display. Participants in the static display condition are prompted to select the district with a specified pickup probability. In contrast, those in the NetHOPs condition must estimate the pickup probability of that district based on the animation. To proceed, participants need to answer the question correctly, though they have multiple attempts available.

Participants engage in a practice trial on the third and fourth instruction pages. The task page of this practice trial includes a unique instructional question, prompting them to select the district they believe most drivers will search in. Participants then proceed to the feedback page, where they receive an explanation of the reward mechanism and learn how their ongoing compensation will be updated and displayed throughout each trial. After completing all 15 trials, participants can describe how they utilized the information display.
\subsubsection{Robustness Checks}
To ensure the robustness of our findings, we conduct two additional replications of our experiment. These replications assess variation in two specific assumptions: trial order and level composition.

\begin{enumerate}
    \item \textbf{Robust Trial Order}: Participants complete trials in a different, yet fixed, order compared to those in the main experiment (see columns 3 and 5 of \autoref{fig:experiment-diagram}). This approach allows us to evaluate to what extent our results might be influenced by the specific trial order.
    \item \textbf{Robust Level Composition}: Participants complete trials in the same order as in the main experiment. However, we alter the level composition by using $\text{Poisson}(3)$ instead of $\text{Poisson}(1.5)$ (refer to column 4 of \autoref{fig:experiment-diagram}). This change results in an increased proportion of more strategically sophisticated L2s within decision scenarios consisting of $N$ drivers. Following the same procedure outlined in Section \ref{sec:level-endownment}, we modify the proportion endowed to L2s from a 40-60 split for L0-L1 (i.e., as in the main experiment) to a 20-80 split for L0-L1. This adjustment allows us to investigate potential variations in results, particularly in how they may manifest when the level distribution leans more towards higher strategic sophistication.
\end{enumerate}

\subsubsection{Participants}
We recruited participants from Prolific using a gender-balanced sample and two pre-screening criteria: fluency in English and being based in the US. Given our staged experimental design and the need to collect data for two additional rounds of robustness checks (i.e., a total of six data collection arms), we excluded participants who had previously taken part in the study, ensuring that each collection arm contained unique Prolific users.

\paragraph{\textbf{Defining Level Composition}} 
We define the distribution over levels for our decision scenarios using a $\text{Poisson}(\lambda)$ where $\lambda = 1.5$ for the main experiment and Robust Trial Order experiment, and a $\lambda = 3$ for the Robust Composition experiment. As described in Section \ref{sec:level-endownment} and demonstrated in \autoref{tab:level-comp-mainExperiment}, we adopt a breakdown of 30\% L0s, 40\% L1s, and 30\% L2s for the main experiment and Robust Trial Order experiment, and a breakdown of 15\% L0s, 35\% L1s, and 50\% L2s for the Robust Composition experiment, following the same procedure.

\paragraph{\textbf{Number of participants to recruit}}\label{sec:no-level1-robustComp}
Given that each decision scenario in our experiment involves hundreds of drivers, we choose to recruit 50\% of L1s and L2s based on the true composition (i.e., see row 5 of \autoref{tab:level-comp-mainExperiment}) for each between-subject treatment of the main experiment (i.e., assuming the minimum number of drivers for a scenario is 598), and 25\% for the two robustness checks. We then re-sample the decisions with replacement from these 50\% and 25\% samples to create the \textit{level-specific outcomes} shown to L2s and the system outcomes, which are generated by combining the decisions of all three levels according to the level distribution defined for each experiment. \autoref{tab:all-collection-sizes} presents the number of participants we recruited by levels and collection arm. Note that the total number of participants we recruit for the Robust Composition experiment is 300. This is because this collection arm follows the same trial order as the main experiment, allowing us to use L1 responses from the main experiment without additional recruiting.

\begin{table*}[ht]
\caption{
The number of participants recruited for each collection arm by size. The parentheses in each table cell include the number of recruited participants for each of the four treatment conditions per level (column) and per collection arm (row).
}
\label{tab:all-collection-sizes}
\begin{threeparttable}

\begin{tabular}{cccc}
\toprule
  Collection       & Level-1 & Level-2 & Total \\
\midrule
Main Experiment    & \begin{tabular}[c]{@{}c@{}}480\\ ($4 \times 120$)\end{tabular} & \begin{tabular}[c]{@{}c@{}}360\\ ($4 \times 90$)\end{tabular} & 840   \\
Robust Trial Order & \begin{tabular}[c]{@{}c@{}}240\\ ($4 \times 60$)\end{tabular}  & \begin{tabular}[c]{@{}c@{}}180\\ ($4 \times 45$)\end{tabular} & 420   \\
Robust Composition & \begin{tabular}[c]{@{}c@{}} 212\\ ($4 \times 53$)\end{tabular}  & \begin{tabular}[c]{@{}c@{}}300\\ ($4 \times 75$)\end{tabular} & 300  \\
\bottomrule
\end{tabular}%
\begin{tablenotes}
\small
\item Note: The 212 L1s from the Robust Composition experiment are sampled from those of the main experiment because L1 tasks are identical for these two collection arms.
\end{tablenotes}   
\end{threeparttable}
\end{table*}

\subsubsection{Analysis Method}
We pre-registered an analysis plan, which includes four dependent variables: two at the aggregate-level and two at the individual-level.

\paragraph{\textbf{Aggregate-level Dependent Variables}} \label{sec:aggregate-response-sampling}
In essence, a principal cares about aggregate-level outcomes, such as the system's efficiency in matching drivers and riders and the discrepancy between predicted and realized outcomes. We compute these aggregate-level outcome variables based on the system outcomes created by integrating decisions from L0 (i.e., derived from taxi data) with those from L1 and L2 (i.e., collected from participants in each trial), using a pseudo-Poisson distribution that determines the frequency distribution across all levels. Since the number of participants we recruit per level per trial is fewer than what is defined by the decision scenario, we sample participants' decisions with replacement to attain the necessary number for each decision scenario. To minimize sampling error, we repeatedly generate system outcomes 500 times per trial, which forms a distribution for each aggregate-level outcome variable. It is important to note that these system outcomes, which differ from the \textit{level-specific outcomes} used to evaluate participants' individual-level decisions, can only be computed after completing both the L1 and L2 stages of data collection.

Given these per-trial outcome distributions, we define the \textbf{social welfare ratio} for each trial as the ratio of the total realized pickups (i.e., the achieved social welfare) to the maximum possible pickups (i.e., the total attainable welfare). We utilize a non-linear optimization via the augmented Lagrange method~\cite{ye1988interior} that employs random initialization and multiple restarts~\cite{hu2009random} to determine the optimal flow distribution for each trial. Taking a flow vector of length three as input, our objective function outputs a numerical value indicative of the number of drivers who failed to secure pickups. This function operates under constraints, ensuring that the input flow vector sums to the total number of drivers in a decision scenario and that the flow to any district stays within historical data bounds. Full details of this process are available in the Supplemental Material.

We calculate the \textbf{distribution shift} for each trial by measuring the Earth Mover's Distance (EMD)~\cite{rubner1997earth, rubner1998metric} between the distribution of deduced search flows displayed to participants and the flows simulated from the system outcomes. EMD quantifies the discrepancy between these two types of flow distributions as the minimum amount of ``work" needed to transform one distribution into the other. To compute EMD, we map both flow distributions on a 2D grid and calculate the minimum Euclidean distance required to shift and align them. This distance is then standardized by the total number of drivers in the decision scenario. We describe the cost function used to optimize the EMD computation in our Supplemental Materials.

\paragraph{\textbf{Individual-level Dependent Variables}}
We define two individual-level response variables to better understand how participants endowed with varying levels of strategic sophistication are affected by our experimental manipulations. The first variable, \textbf{best response}, is a binary indicator reflecting whether a participant selected the district with the highest predicted pickup probability, as determined by the endowed level. For L1s, the best response amounts to simply choosing the district with the highest predicted pickup probability, as shown on the display. For L2s, it involves selecting the district yielding the highest predicted pickup probability, based on the \textit{level-specific outcome} derived from combining L0 and L1s' decisions under the endowed proportion of L2s. Evaluating the best responses of L1s and L2s over repeated decisions allows us to observe the impact of showing prediction uncertainty and realized prediction error. Specifically, it reveals how these factors influence the decision-making of agents who operate under various assumptions about the behavior of others.

\textbf{Anticipation error} measures how well a participant anticipates other participants' search decisions. Evaluating anticipation error over repeated decisions allows us to observe how the provision of realized prediction error, particularly in cases where the displayed predictions are ``inaccurate", aids in anticipating the behavior of others. Similar to the calculation of the distribution shift, we compute the EMD between each participant's anticipated flow distributions (i.e., their anticipation of how many drivers would choose each district) and the \textit{level-specific outcome} used to evaluate decisions under their endowed belief (as detailed in Section \ref{sec:level-specific-feedback}).

\paragraph{\textbf{Statistical Models}}
We pre-registered regression models for welfare ratio and distribution shift to analyze the two aggregate-level variables, using the expected values derived from 500 simulations as the dependent variables. We specified the maximal models~\cite{barr2013random} for the aggregate-level variables by Wilkinson-Rogers-Pinheiro-Bates syntax~\cite{pinheiro2017package, wilkinson1973symbolic} as follows:

\begin{minipage}[t]{0.46\textwidth}
\begin{agmodel}[H]
    \centering
    \caption{Welfare Ratio (\textit{ratio})}\label{mod-ratio}
    \begin{algorithmic}[1]
        \State \textit{ratio} $\sim$ Beta ($\mu$, $\phi$) \label{lst:line:ratio-likelihood}
        \State logit($\mu$) = \textit{display} * \textit{feedback} * \textit{trial} \label{lst:line:ratio-mu-linear-model}
        \State log($\phi$) = \textit{display} * \textit{feedback} * \textit{trial} + \textit{maxProb} \label{lst:line:ratio-phi-linear-model}
    \end{algorithmic}
\end{agmodel}
\end{minipage}
\hfill
\begin{minipage}[t]{0.46\textwidth}
\begin{agmodel}[H]
    \centering
    \caption{Distribution Shift (\textit{DS})}\label{mod-ds}
    \begin{algorithmic}[1]
        \State \textit{log(DS)} $\sim$ student\char`_t ($\nu$, $\mu$, $\sigma$) \label{lst:line:ds-likelihood}
        \State $\mu$ = \textit{display} * \textit{feedback} * \textit{trial} \label{lst:line:ds-linear-model}
    \end{algorithmic}
\end{agmodel}
\end{minipage}
\bigskip

Here, \textit{trial} and \textit{maxProb} are numeric, while \textit{display} and \textit{feedback} are categorical. For the welfare ratio model, we use a Beta distribution as the likelihood (line 1, left), parameterized by mean ($\mu$) and precision ($\phi$). We apply a logit link for $\mu$ (line 2, left) and include all experimental variables, \textit{display}, \textit{feedback}, and \textit{trial}, as predictors on $\mu$ of the likelihood. We apply a log link for $\phi$ (line 3, left) and include an additional predictor \textit{maxProb}, which is the maximum pickups optimized based on the system outcome to control the amount of spread for the achieved welfare. For the distribution shift model, we use a student-t distribution as the likelihood for log-transformed \textit{DS} (line 1, right), parameterized by $\nu$ (degree of freedom), $\mu$ (mean), and $\sigma$ (scale) and include all experimental variables as predictors for $\mu$ (line 2, right).

To model individual-level variables while accounting for the experimental design, we pre-registered Bayesian Linear Mixed-effect Models to analyze the best response and anticipation error. We specified separate models by levels since the tasks that an L1 completes are different from those completed by an L2, and separate models by collection arms since the level composition or trial order can change for the two robustness checks. We specify the individual-level maximal models as follows:

\begin{minipage}[t]{0.46\textwidth}
\begin{indmodel}[H]
    \centering
    \caption{Best Response (\textit{BR})}\label{mod-br}
    \begin{algorithmic}[1]
        \State \textit{BR} $\sim$ Bernoulli ($p$) \label{lst:line:br-likelihood}
        \State logit($p$) = \textit{display} * \textit{feedback} * \textit{trial} + (\textit{trial} | \textit{ID})\label{lst:line:br-linear-model}
    \end{algorithmic}
\end{indmodel}
\end{minipage}
\hfill
\begin{minipage}[t]{0.46\textwidth}
\begin{indmodel}[H]
    \centering
    \caption{Anticipation Error (\textit{AE})}\label{mod-ae}
    \begin{algorithmic}[1]
        \State \textit{log(AE)} $\sim$ student\char`_t ($\nu$, $\mu$, $\sigma$) \label{lst:line:ae-likelihood}
        \State $\mu$ = \textit{display} * \textit{feedback} * \textit{trial} + (\textit{trial} | \textit{ID}) \label{lst:line:ae-linear-model}
    \end{algorithmic}
\end{indmodel}
\end{minipage}
\bigskip

Here, \textit{trial} is numeric, while \textit{display}, \textit{feedback}, and \textit{ID} are categorical, with \textit{ID} representing each participant's unique Prolific ID. Line 1 of both models presents the assumed likelihood functions. For the binary best response, we utilize a Bernoulli distribution as the likelihood with parameter $p$ to represent the best response rate. To handle potential outliers in participants' anticipation, we apply a Student-t distribution as the likelihood for log-transformed \textit{AE}, parameterized by $\nu$ (degree of freedom), $\mu$ (mean), and $\sigma$ (scale). Line 2 presents the hierarchical linear models, for which we use a logit link for the best response rate $p$. Both individual-level models estimate fixed effects of trial, display type (i.e., static or NetHOPs), and feedback structure (i.e., bandit or full). Because participants may have varying baseline performances and demonstrate different variations in performance across trials, we incorporate random effects for each participant, including random intercepts and slopes for the effects of \textit{trial} and the intercepts. 

We adhere to a standard Bayesian workflow~\cite{gelman2020bayesian} to check our model fits and include model diagnostics in the Supplemental Material. We report expected performance with the median point estimate of the expectation with uncertainty expressed as a 95\% highest posterior density interval (i.e., credible interval, CI) for each variable and condition, marginalizing over trials unless examining learning effects. 
\section{Results}
In total, we received 1,573 valid responses. \autoref{tab:experiment-recruitment} summarizes the number of unique participants by treatment, level, and collection arm, with these numbers highlighted in bold. We excluded four participants from the total number recruited according to our pre-registered exclusion criteria (i.e., two L1s from the main experiment and two L2s from the Robust Trial Order experiment).

\begin{table*}[ht]
\begin{threeparttable}
\caption{The number of valid responses received by condition, level, and collection arm.}
\label{tab:experiment-recruitment}
\begin{tabular}{ccccccc}
\toprule
Collection & \multicolumn{2}{c}{Main Experiment} & \multicolumn{2}{c}{Robust Trial Order} & \multicolumn{2}{c}{Robust Composition} \\
Level-$k$     & Level-1    & Level-2     & Level-1    & Level-2    & Level-1    & Level-2     \\
\midrule
Valid       & \textbf{480}        & \textbf{361}         & \textbf{240}        & \textbf{181}        & 212        & \textbf{311}         \\
\midrule
Static+Bandit & 115 & 103 & 63 & 40 & 53 & 78 \\
Static+Full & 115 & 86 & 59 & 44 & 53 & 70 \\
NetHOPs+Bandit & 122 & 92 & 52 & 49 & 53 & 84 \\
NetHOPs+Full & 128 & 80 & 66 & 48 & 53 & 79 \\
\bottomrule
\end{tabular}%
\begin{tablenotes}
\small
\item Note: The 212 L1 responses from the Robust Composition experiment are sampled from those of the main experiment because the L1 tasks are identical for these two collection arms.
\end{tablenotes}
\end{threeparttable}
\end{table*}

\subsection{Data Preliminaries}
\autoref{tab:completion-time-by-tx} presents the study completion times, from which we see that the completion time depends on the participants' endowed levels, the type of display they viewed, and the feedback structure they received. Notably, participants with L2 beliefs, who interacted with the NetHOPs display, and who received full feedback generally took longer to complete the study.

\begin{table*}[ht]
    \caption{Summary of study completion time by treatment condition.}
    \label{tab:completion-time-by-tx}
    \begin{tabular}{cccccc}
        \toprule
        Group & Min. & Median & Mean & SD & Max \\ 
        \midrule
        Pooled & 4.4 & 18.1 & 20.4 & 10.6 & 132 \\
        Level-1 & 4.4 & 17.4 & 19.4 & 9.5 & 77\\
        Level-2 & 5.1 & 18.9 & 21.5 & 11.5 & 132\\
        \midrule
        Static + Bandit & 4.4 & 16.8 & 18.8 & 10 & 83.1 \\
        Static + Full & 5.9 & 18.7 & 21.1 & 10.8 & 78.8 \\
        NetHOPs + Bandit & 5.1 & 16.4 & 19.3 & 10.9 & 132 \\
        NetHOPs + Full & 6.2 & 20.7 & 22.5 & 10.2 & 77 \\
        \bottomrule
    \end{tabular}
\end{table*}

Recall our participants could optionally describe how they used the interface to make decisions. We lightly analyzed their statements by sampling 100 comments from each level across collection arms. From 84 meaningful L1 comments, we identified four participants who either misunderstood the payoff information shown on the display or misinterpreted the strategic setting. Similarly, of the 83 meaningful L2 comments, we observed that the level-endowment success rate is slightly lower (84\%), with 13 participants acting like L1s or randomly guessing. These findings suggest that the actual Poisson distribution realized in our experiments may be more skewed towards L0s and L1s than we originally intended. 

\subsection{Aggregate-level Outcome}
\begin{figure}[ht]
  \centering
  \includegraphics[width=\linewidth]{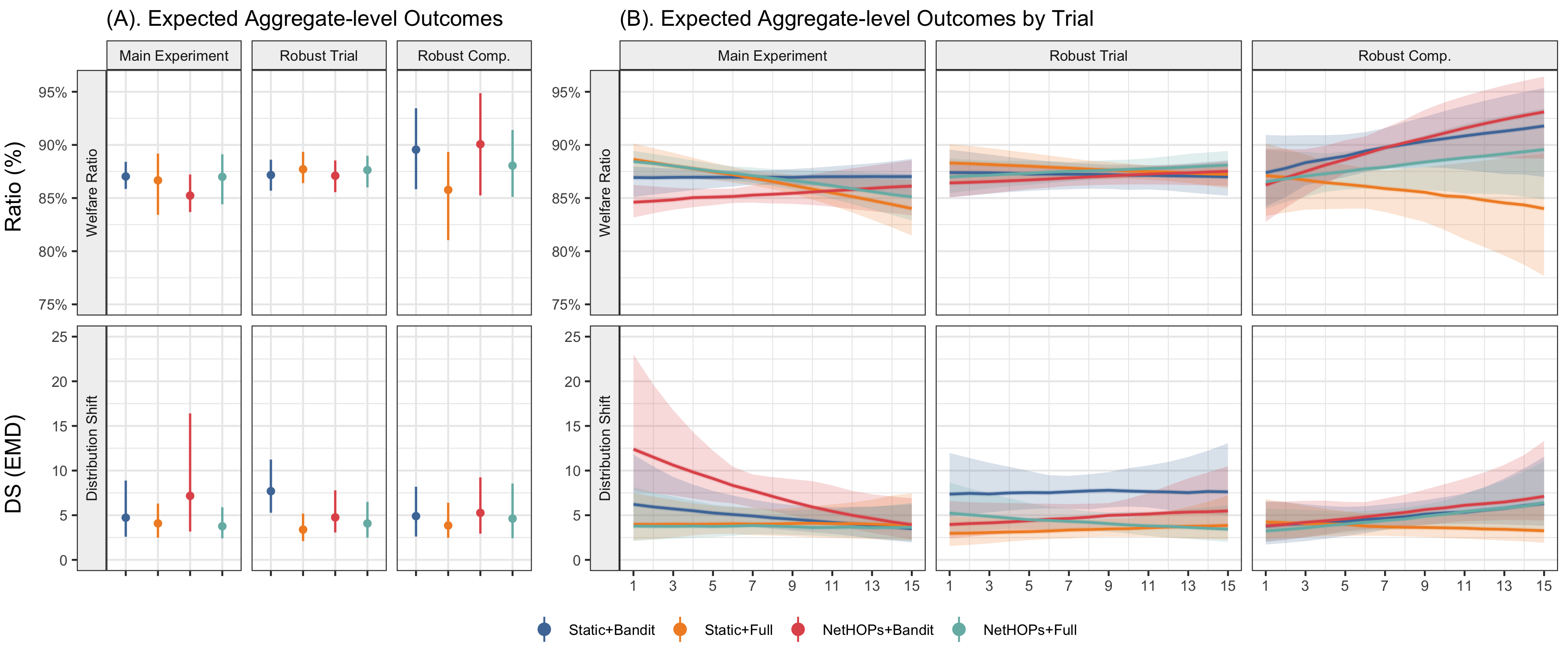}
  \caption{(A) Median point estimates of the expected welfare ratio (top) and distribution shift (bottom) resulted from the system outcome. We expressed uncertainty as 95\% credible intervals (CIs) predicted by the fixed effects of both aggregate-level models for each treatment marginalized over trials. (B) Same as (A), we present both the point estimate and the corresponding 95\% CIs for the welfare ratio (top) and distribution shift (bottom) for each trial to examine changes in system outcome by trial.}
  \label{fig:aggregate-results}
  \Description{Aggregated Results}
\end{figure}

We first examine the two aggregate-level variables, the welfare ratio and the distribution shift, derived from the system outcomes by combining the decisions according to the true mixture of levels. The welfare ratio in our experimental context represents system efficiency, measured as the proportion of realized pickups achieved relative to the maximum possible in a given decision scenario. The distribution shift quantifies the difference, in EMD, between the deduced flow displayed to the participants and the realized flow of the system. We present the 95\% CIs of the median expected outcomes from the posterior predictive distribution across different treatments in \autoref{fig:aggregate-results} (A), and examine their variations by trial in \autoref{fig:aggregate-results} (B). 

\subsubsection{Welfare Ratio}
\paragraph{Welfare Ratio by Interface}
When using a $\text{Poisson}(1.5)$ to define a 30-40-30 mixture of levels (i.e., the main experiment and the Robust Trial Order experiment), we observe similar expected welfare ratios across treatment interfaces. The expected ratios consistently range between 83\% and 89\%. (\autoref{fig:aggregate-results} (A) top-left and top-middle). However, in the Robust Composition experiment, which utilizes a $\text{Poisson}(3)$ leading to a more skewed split toward higher-levels (15-35-50), we note a higher median expected ratio on average with greater variance across interfaces. As for the display type, whether NetHOPs or static displays, it appears to minimally affect the welfare ratio of the system outcomes. Nevertheless, bandit feedback (90\%; CI[86\%, 94\%]) tends to yield slightly higher welfare ratios compared to full feedback (87\%; CI[82\%, 91\%]). 

\paragraph{Trial-level Variations}
We observe consistent interface effects on the expected welfare ratio by trial, as detailed in row 1 of \autoref{fig:aggregate-results} (B). In the main experiment, changes in the welfare ratio over trials seem predominantly influenced by the feedback structure. Specifically, bandit feedback is associated with a slight increase in the welfare ratio across trials (red and blue), whereas full feedback shows a slight decrease (yellow and green). In the Robust Trial Order experiment, the trial-level variation in welfare ratios is similar to that observed from the main experiment but with a less pronounced feedback effect, as indicated by the flatter slopes. In scenarios with a level distribution that skews more toward higher levels, as seen in the Robust Composition experiment, welfare ratios appear similar in early trials but gradually diverge. Notably, all interfaces exhibit an increase in welfare over time, except for static displays with full feedback (yellow), where we notice a decrease in welfare. 

In summary, we observe relatively robust welfare variations across the three experiments. These variations are particularly evident in (1) static displays and full feedback, which tend to result in a lower welfare ratio by trial, and (2) NetHOPs displays and bandit feedback, which generally lead to a higher welfare ratio over trials. 

\subsubsection{Distribution Shift}

\paragraph{Distribution Shift by Interface}
Under a 30-40-30 mixture of levels from a $\text{Poisson}(1.5)$, as used in both the main experiment and the Robust Trial Order experiment, our results indicate that the full feedback typically leads to lower and less extreme expected distribution shift compared to bandit feedback on average (e.g., in the main experiment, Full 3.9; CI[2.4, 6.3] and Bandit 5.7; CI[2.6, 14.1]). This trend is particularly pronounced with NetHOPs displays, where the variance in the expected distribution shift under bandit feedback is higher, as illustrated in \autoref{fig:aggregate-results} (A) bottom-left (red). We see a smaller version of this effect in the Robust Composition experiment. Static displays may produce slightly lower and less extreme expected distribution shift than NetHOPs (e.g., main experiment, Static: 4.4; CI[2.5, 8.1] and NetHOPs: 4.7; CI[2.5, 14.1]), but these estimates have high uncertainty. 

\paragraph{Trial-level Variations}
In the main experiment, as depicted in \autoref{fig:aggregate-results} (B) bottom-left, changes in distribution shift over trials appear to be influenced by the feedback structure. With bandit feedback, the distribution shift decreases as trials progress. With full feedback, the distribution shift remains relatively consistent across trials. Notably, despite substantial variation in distribution shift during initial trials by interface, the expected distribution shift converges by the final trial across all interfaces. In contrast, the Robust Trial Order experiment (\autoref{fig:aggregate-results} (B) bottom-middle) presents a contrasting pattern, where the changes in distribution shift over trials are more similar, as indicated by the flatter slopes. In the Robust Composition experiment (\autoref{fig:aggregate-results} (B) bottom-right), the expected distribution shift generally increases with each trial across interfaces, except in the case of static displays paired with full feedback (yellow). Although the patterns of distribution shift vary depending on trial order and level composition, static displays with full feedback consistently result in a slightly lower distribution shift across the experiments. 

\subsection{Individual-level Analysis}
We analyzed participants' individual-level best response rate and anticipation error to investigate the underlying causes of the variations observed in the aggregate-level variables. Recall that in our experimental design, a \textit{best response} implies a participant's decision to search a district with the highest pickup probability, as determined by the level-specific outcome. Meanwhile, \textit{anticipation error} quantifies the difference, in EMD, between the anticipated flow and the actual flow of the level-specific outcome. We present both individual-level response variables using the 95\% CIs of the median expected values from the posterior predictive distribution by treatment conditions and examine them by trial.

\begin{figure}[ht]
  \centering
  \includegraphics[width=\linewidth]{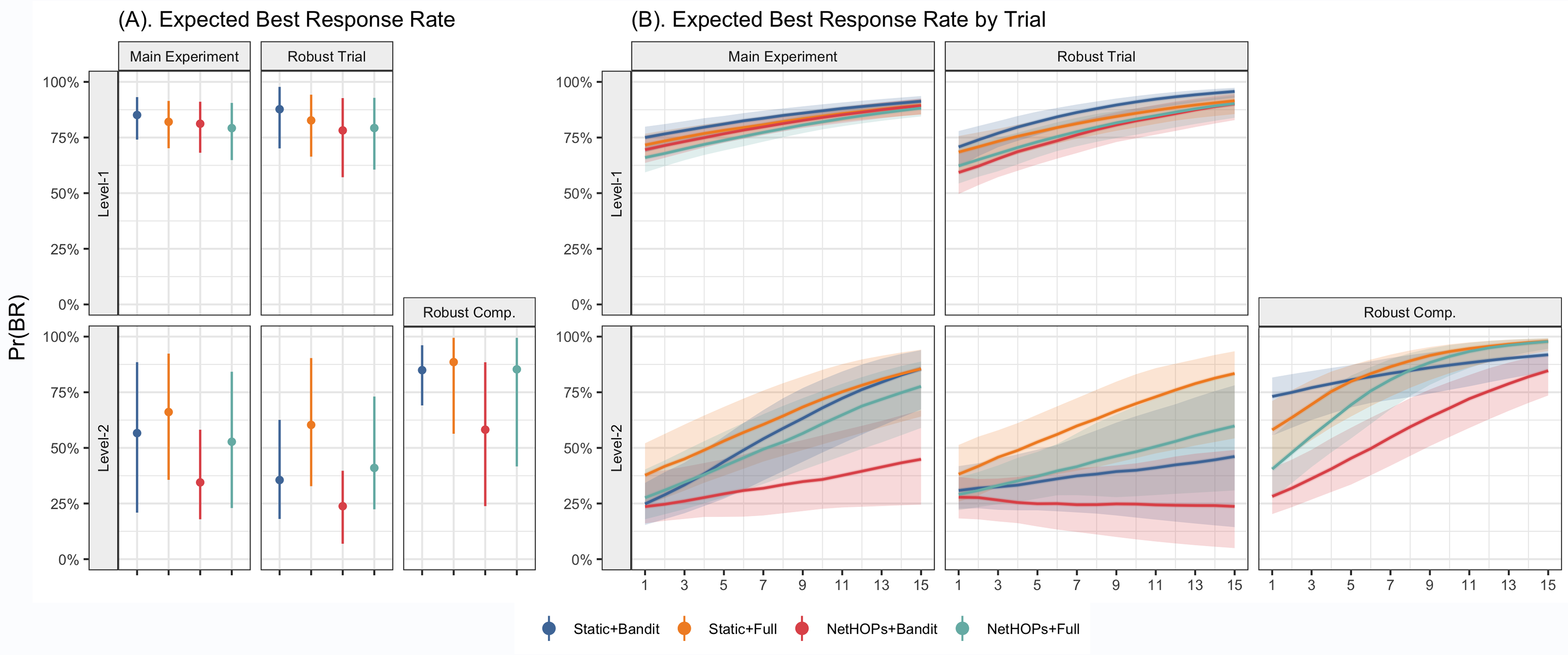}
  \caption{(A) The median point estimates of the expected best response rate with uncertainty expressed as 95\% credible intervals (CIs) predicted by the fixed effects and marginalizing over trials. (B) As in (A), we present both the point estimate and the corresponding 95\% CIs of the expected BR rate for each trial to examine changes by trial. \textit{\textbf{Notice we omit L1s from the Robust Composition experiment because these L1s' responses are sampled from those of the main experiment.}}}
  \label{fig:brRate-results}
  \Description{Results of BR rate}
\end{figure}

\subsubsection{Level-1 Best Response}
We observe minimal differences in L1s' best response rate across treatments in both the main experiment and the Robust Trial Order experiment, as shown in the top-left and top-middle sections of \autoref{fig:brRate-results} (A). High best response rate among L1s are expected, given their endowed belief in the ``accuracy" of the displayed predictions; their best response involves choosing the district with the highest predicted pickup probability shown on the display. In both experiments, we find that (1) static displays lead to slightly higher best response rate for L1s compared to NetHOPs (e.g., in the main experiment, Static: 84\%; CI[71\%, 93\%] and NetHOPs: 80\%; CI[66\%, 91\%]), and (2) bandit feedback slightly outperforms full feedback (e.g., in the main experiment, Bandit: 83\%; CI[70\%, 93\%] and Full: 80\%; CI[67\%, 92\%]). This indicates that when responding to straightforward displays, processing prediction uncertainty and viewing full feedback might lead some participants to second-guess their best response. However, these interface effects on L1s are relatively small. 

In \autoref{fig:brRate-results} (B) row 1, we observe slight but consistent improvements in L1s' best response rate across all treatments as they completed more trials in both the main experiment and the Robust Trial Order experiment. It appears that L1s using static displays and bandit feedback (blue) are more likely to best respond than those using alternative interfaces. 

\subsubsection{Level-2 Best Response}
L2s' best response rate, as illustrated in \autoref{fig:brRate-results} (A) row 2, are much lower than those of L1s across treatments, exhibiting greater uncertainty in their estimates. This lower rate among L2s is likely attributed to the more challenging task they face: they need to reason about the distribution of other players' actions using a display that is not inherently ``accurate". The larger variance that we observe is partly due to high trial-level variance, suggesting that some L2s were able to considerably improve their best response rate through repeated decisions. 

At a high level, our findings indicate that (1) L2s who receive full feedback are slightly more inclined to best respond compared to those with bandit feedback (e.g., in the main experiment: Full 59\%; CI[26\%, 90\%] and Bandit 40\%; CI[19\%, 85\%]), and (2) L2s viewing static displays are more likely to best respond than those viewing NetHOPs (e.g., in the main experiment: Static 61\%; CI[25\%, 91\%] and NetHOPs 40\%; CI[18\%, 80\%]). These effects persist in the Robust Trial Order and Robust Composition experiments, though with overlapping intervals. The use of static displays and full feedback appears to be the most effective in helping L2s to best respond. Our results provide weak evidence that the emphasis on prediction uncertainty could complicate L2s' anticipation of others' responses to a display, as speculated by \citet{kayongo2021visualization} but not empirically validated in their study. However, the high uncertainty in the estimated L2 performance, likely due to the challenge of best responding to a display under beliefs about a mixture of other agents over levels, prevents us from drawing any strong conclusions.

Another difference we observe is that the best response rate for L2s are noticeably higher across treatments in the Robust Composition experiment. We expect this is because the presence of more L1s in the mixture of levels that L2s are endowed with (i.e., 80\% L1s in the Robust Composition experiment vs. 60\% L1s in the main experiment) may make it easier for L2s to anticipate others' actions. With a greater number of lower-level players in the mixture that L2s are endowed with, more competitors would tend to search in the most lucrative district following the display. 

When examining L2s' best response rate by trials, as shown in \autoref{fig:brRate-results} (B) row 2, we find that L2s exhibit steeper increasing slopes compared to L1s, suggesting that, with more trial completions, L2s are capable of learning to map between predicted and level-specific outcomes. We see consistent treatment effects where static displays and full feedback tend to help L2s best respond over trials. We note that L2s using interfaces less suited for their tasks, namely NetHOPs and bandit feedback, may cease to learn at all and can be influenced by the order of decision scenarios, as seen in the Robust Trial Order experiment in \autoref{fig:brRate-results}, bottom-middle. 

\begin{figure}[ht]
  \centering
  \includegraphics[width=\linewidth]{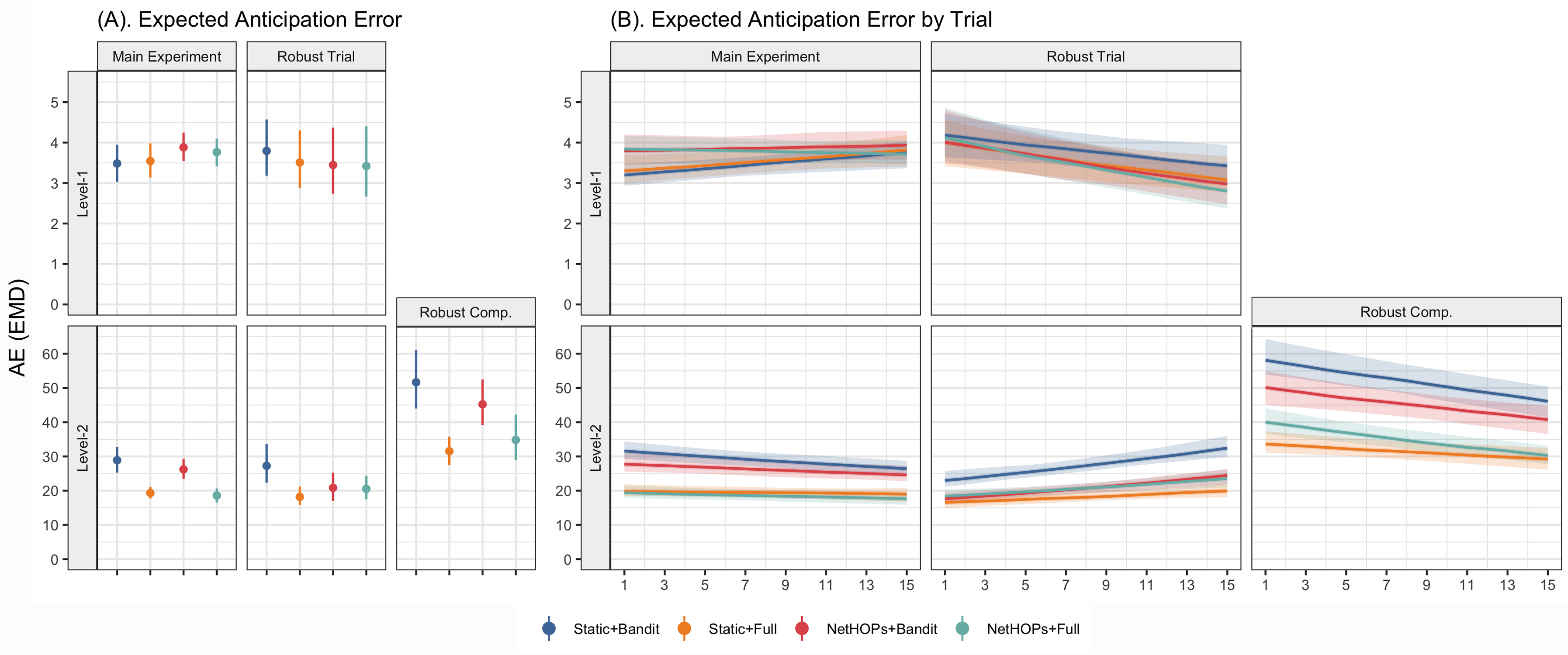}
  \caption{(A) Median point estimates of the expected anticipation error in EMD with uncertainty expressed as 95\% credible intervals (CIs), accounting for all fixed effects and marginalizing over trials. EMD describes the averaged Euclidean distance a participant's anticipated flow distribution must move in order to match the flows of the level-specific outcome.  (B) Same as (A), we present both the point estimate and the corresponding 95\% CIs of the expected anticipation error for each trial to examine changes in performance by trial. \textbf{\textit{Notice the y-axis scales are different between L1s (top) and L2s (bottom), and we omit L1s from the Robust Composition experiment because these L1s' responses are sampled from those of the main experiment}}.}
  \label{fig:ae-results}
  \Description{Results of anticipation error}
\end{figure}

\subsubsection{Level-1 Anticipation Error}
Between our main experiment and the Robust Trial Order experiment, the anticipation error for L1s, measured in Euclidean distance using EMD, is quite small, with narrowly overlapping CIs ranging between 2.7 and 4.5, as shown in \autoref{fig:ae-results} (A) row 1. When examining the variation in L1s' anticipation error over trials presented in \autoref{fig:ae-results} (B) row 1, although the signs of the slopes vary, all slopes are small. This indicates minimal improvement or deterioration in L1s' ability to anticipate level-specific outcomes.

\subsubsection{Level-2 Anticipation Error}
For L2s, the expected anticipation error is much higher across all treatments compared to that of L1s. As demonstrated in \autoref{fig:ae-results} (A) row 2, we observe strong feedback effects across all three experiments. L2s who receive full feedback have a clear advantage in anticipating competitors' actions over those who receive bandit feedback (e.g., in the main experiment, Full: 19; CI[16.9, 21] and Bandit: 27.3; CI[23.8, 32.2]). We observe no distinct effect when comparing the use of NetHOPs versus static displays. 

We observe a higher anticipation error for L2s in the Robust Composition experiment. This may seem contradictory to the results of L2s' best response rate observed from the same experiment, which is higher compared to the other experiments, presumably because of the increased proportion of L1s they are best responding to. We speculate that anticipation error is a more nuanced measure. For L2s, precisely estimating how many other competitors will choose to search in each district presents a greater challenge, even if identifying the best response is relatively straightforward. 

Examining L2s' anticipation error by trial, as shown in \autoref{fig:ae-results} (B) row 2, indicates that full feedback consistently aids participants in more accurately anticipating others' decisions across trials in all three experiments. The results from the Robust Trial Order experiment suggest that the variations in anticipation error over repeated decisions can depend on the specific sequence of prior decision scenarios. For instance, anticipation error decreases in the main experiment and the Robust Composition experiment, while it increases in the Robust Trial Order experiment. 

\begin{figure}[ht]
  \centering
  \includegraphics[width=\linewidth]{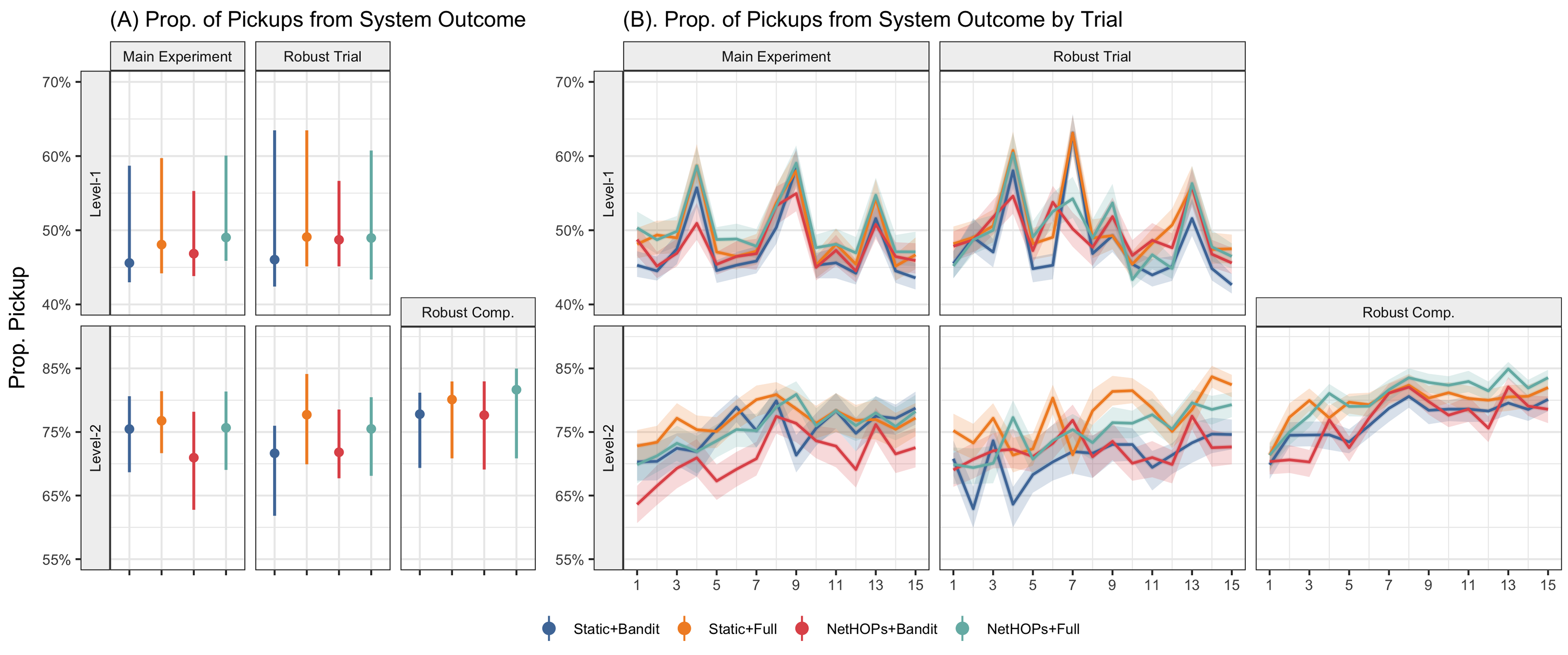}
  \caption{
  (A) The median proportion of L1s and L2s who secured pickups from their selected district with 95th percentile intervals (PI) summarizes the distribution of pickup proportions from 500 simulations of the system outcomes, marginalized over trials. (B) As in (A), we present both the median and 95\% PI of L1 and L2s' pickup proportion for each trial to examine changes by trial. \textbf{\textit{Notice the y-axis scales are different between L1s (top) and L2s (bottom), and we omit L1s from the Robust Composition experiment because these L1s' responses are sampled from those of the main experiment}}.
  }
  \label{fig:prop-system-outcome}
  \Description{Results of anticipation error}
\end{figure}

\subsection{Connecting Individual-level Responses and Aggregate-level System Outcomes}
In our analysis of the individual-level best response rate and anticipation error against level-specific outcomes, a question naturally arises: Do similar display or feedback effects manifest when evaluating participants' decisions against the system outcome realized by combining decisions from L0s, L1s, and L2s? To address this, we construct 95th percentile intervals (PIs) using the median as the point estimate. This method assesses the proportion of L1s and L2s who successfully secured pickups from their selected district based on 500 simulations of the system outcomes for each type of interface. These outcomes are marginalized over all trials, as depicted in \autoref{fig:prop-system-outcome} (A), and analyzed by individual trials in \autoref{fig:prop-system-outcome} (B).

From our simulations, the PIs for L1s' proportion of pickups, as shown in \autoref{fig:prop-system-outcome} (A) row 1, indicate a strong right-skewness. Most simulated observations fall below or close to 50\%, suggesting that, in general, the likelihood of pickups for L1s is relatively low and associated with high uncertainty. When marginalizing the proportion of pickups for L1s across all simulations in the three experiments, the effects of display type and feedback structure for L1s appear to be minimal. Specifically, the difference in proportion between Static display and NetHOPs is -0.6\% (PI[-6\%, 7\%]), while that between full and bandit feedback is 1.7\% (PI[-4\%, 7\%]). 

The noticeable difference between the proportions of pickups for L1s and L2s, as illustrated in row 1 and row 2 of \autoref{fig:prop-system-outcome} (A), is that L2s have a clearly higher chance of securing pickups due to their strategic sophistication. Similar to the results when evaluating L2s' best response rate under their level-specific outcomes, we find minimal differences in pickup probability for L2s when using a static display versus a NetHOPs display in the system outcome, with a difference of 0.8\% (PI[-6.9\%, 8.6\%]). We again observe higher differences in the pickup probability when L2s use full feedback, amounting to 3.4\% (PI[-3\%, 10.4\%]). In summary, the interface effects identified from our individual-level results appear to be consistent when evaluating decision-making against the system outcome, as reflected by each level's proportion of pickups.

\subsubsection{Level-$k$ Decisions and Welfare Ratio}
The observed trial-level variations in the social welfare ratio can be contextualized by examining several aspects of individual-level decision-making. Design choices, such as displaying prediction uncertainty and error, have a limited impact on L1s who respond to an ``accurate" display. Given the simplicity of their tasks, L1s can easily identify the district with the highest predicted pickup probability (e.g., West) as their best response. In contrast, L2s who face more challenging tasks are able to improve their decision-making when using interfaces with static displays and full feedback. Although the best response rate of L2s is generally lower than that of L1s, as some L2s may initially behave similarly to L1s due to a lack of comprehension of their endowed level, the provision of realized prediction error (i.e., full feedback) can assist L2s in identifying their best response district (e.g., East) over repeated decisions. Consequently, as more trials are completed, an increasing number of L2s, who initially misunderstood their level endowment and behaved like L1s, begin to recognize the opportunity to profitably deviate from the L1s' best response district.

It is important to note that each district in our congestion game has a limited capacity for pickups (see \autoref{fig:modelFits-appendix} in Appendix \ref{sec:counterfactual-appendix}). While L2s deviating to their best response district can increase their overall share of pickups, this simultaneously decreases the probability of successful pickups. Consequently, the remaining district (e.g., North), which is not the best response for either L1s or L2s, emerges as a new opportunity for profitable deviation. However, this opportunity often goes unnoticed by both L1s and L2s, who are myopic, leading to a reduction in the social welfare ratio achieved by the system outcome. This suggests that design choices that support individual-level decisions for certain groups can lead to a worse collective system outcome.

\subsubsection{Level-$k$ Decisions and Distribution Shift}
The use of EMD for quantifying distribution shift effectively describes the overall difference between the flows displayed and those observed in the system outcome. However, the EMD computation does not provide details on the flow differences at the district level, particularly the disparity between the deduced number and the actual number of drivers searching in each district. Since individual-level decisions primarily drive flows to each district, the extent of distribution shift that arises from a system outcome is intrinsically associated with the number of participants who can successfully select the best response strategy based on their endowed levels. Furthermore, we observe that interface design factors can influence distribution shift. Specifically, interfaces incorporating full feedback have a robust feedback effect, resulting in system outcomes with lower distribution shift compared to those with bandit feedback. To further explore this phenomenon, we conducted an exploratory analysis examining trial-level variations in distribution shift, focusing on the flow proportion to each district under different feedback structures. Our objective is to answer the high-level question: How does the feedback structure influence the flows to each district, thereby affecting distribution shift over trials?

\begin{figure}[ht]
  \centering
  \includegraphics[width=.7\linewidth, keepaspectratio]{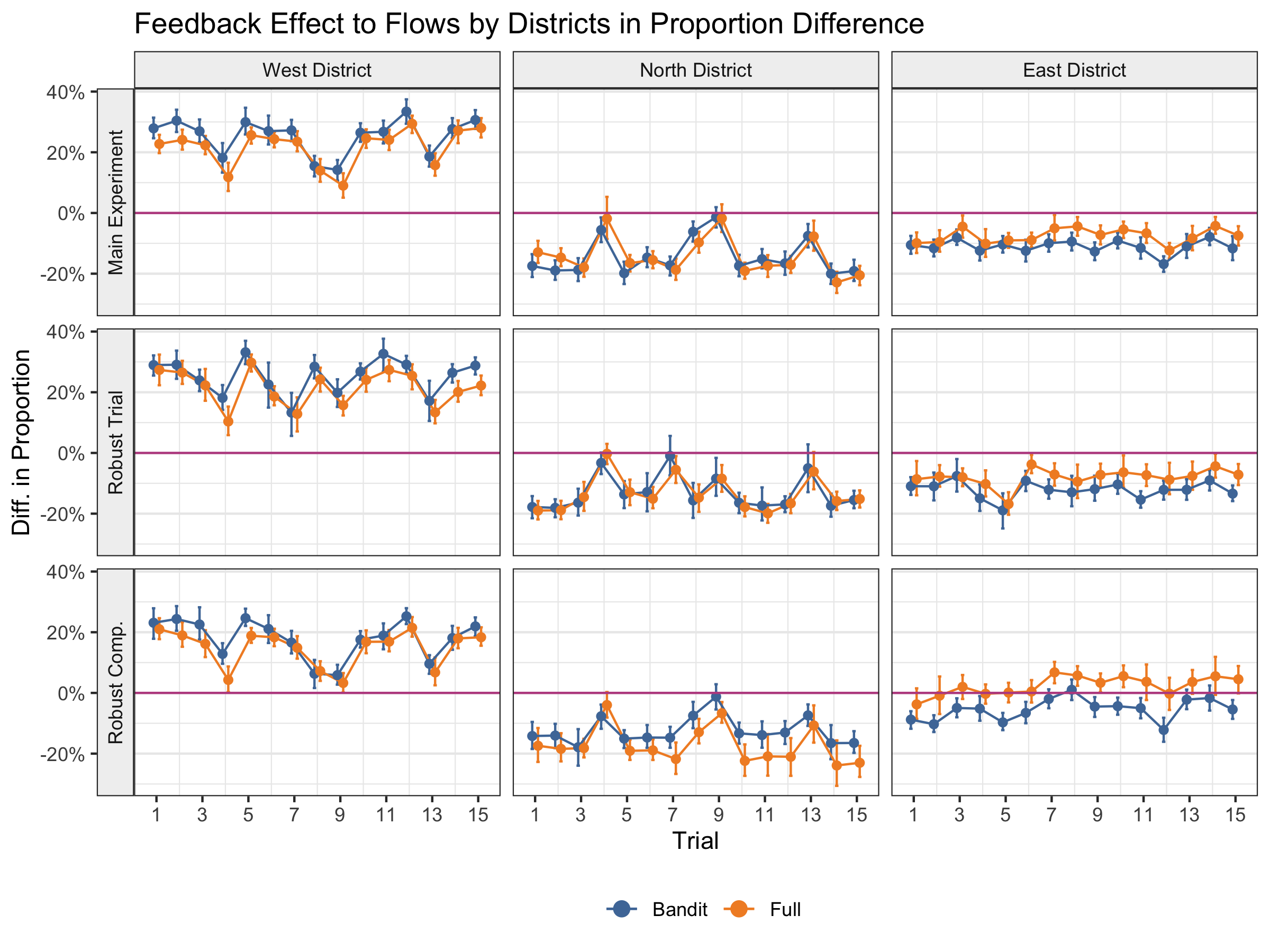}
  \caption{ 
  The median proportion of drivers who searched in each district relative to the deduced flow shown in the information display. The intervals representing the 95th percentile summarize district flows from 500 simulations. The red horizontal line centered at 0 can be viewed as a benchmark, such that a positive proportion difference means more drivers go to a district, while a negative proportion difference means fewer drivers go to a district. Colors differentiate interfaces with full feedback from bandit feedback.
}
  \label{fig:ds-flow}
  \Description{DS with Flow}
\end{figure}

In \autoref{fig:ds-flow}, we dissect the flows of the system outcome for each trial by district (i.e., West, North, and East) for each collection arm. We then present the differences between the observed flows for each district relative to those shown on the information display. As system outcomes are generated by simulations, we create 95th PIs to summarize the variation in proportions going into each district across the simulations. Inspecting the differences in the proportion corresponding to each district across trials reveals a consistent pattern: more participants opt to search the West district (i.e., above the red horizontal line centered at $0\%$ in \autoref{fig:ds-flow}, column 1), while fewer participants choose the East or North districts, compared with the flows shown on the display. This behavior aligns with our expectations: (1) choosing the West District is the best response strategy under L1s' endowed beliefs (see \autoref{tab:trial-orders} of Appendix \ref{sec:trial-BR}), and (2) level endowment may fail for some L2s, who instead choose to search in the West district because it is the most lucrative location shown on the display.

When inspecting the influence of receiving full feedback (i.e., yellow intervals in \autoref{fig:ds-flow}) on the difference in flow proportion to each district, we observe that relatively more participants, including both L1s and L2s, choose to search in the East district (L2s' best response strategy), while fewer search in the West district (L1s' best response strategy). Recall our individual-level results on best response rate suggest that (1) L1s tend to have relatively high and increasing best response rate across trials, and (2) full feedback assists L2s with best responding. Therefore, the gaps representing the difference in flow proportions by feedback structure observed from the West and East districts suggest that as participants complete more trials, more L2s are able to identify their best response strategy corresponding to their endowed level (see \autoref{fig:brRate-results} (B), row 2, yellow). As more L2s choose to profitably deviate and L1s maintain a high best response rate, the distribution shift of the system outcome may reduce. Hence, our results imply that a principal can expect reduced distribution shift by tailoring the interface to the needs of more strategically sophisticated players within a repeated strategic decision-making context. As the amount of distribution shift decreases with repeated decisions, predictions shown on the information display become retrospectively more ``accurate" relative to the system outcome, thereby improving trust and reliance on the information display.
\section{Discussion}\label{sec:discussion}
The results of our experiments underscore the impact of design choices, particularly the provision of prediction uncertainty and error, on the decision-making processes of agents with varying levels of strategic sophistication. Methodologically, our study demonstrates the utility of level endowment and behavioral models in unraveling the complex dynamics of individual-level decision-making and aggregate-level system outcomes. We introduce a staged experimental design as an alternative to synchronous experiments. Moreover, our work highlights the importance of Cognitive Hierarchy Models in the design of information displays for strategic settings, aligning with calls for behavioral researchers to take effect heterogeneity more seriously~\cite{bryan2021behavioural,gelman2023causal}. 

More specifically, our results suggest that the more strategic the user population of the prediction is, the more crucial it becomes to design information displays effectively for anticipating how other agents will respond to the predictions. According to our findings, one way to facilitate this anticipation is to use a static display rather than emphasizing uncertainty. More importantly, when the prediction of an information display may seem inaccurate in hindsight due to agents' strategic responses, providing information on the realized prediction error can aid agents in making more informed decisions despite distribution shift. This result has implications for the study of trust in data-driven predictions. Prior work~\cite{lu2021human} suggests that users' trust in predictive model outputs is affected by the observed accuracy of the model. In real-world scenarios, decision-makers perceiving a gap between predicted and realized outcomes might lose trust and stop relying on the predictions instead of trying to infer how they are wrong and adjusting their decision strategies. However, in a strategic setting like the one we study, a consistently ``wrong'' information display that is transparent about its error may prove more useful to agents than one that is not forthright about its error. An interesting pursuit for future work is to explore agents' trust in such settings. For example, when access to an information display or post hoc decision feedback carries a cost, it raises the question of what factors influence agents' willingness to tolerate an inaccurate display that can potentially improve their decision-making and how agents can effectively assess performance gains in the presence of an incorrect display.

Another important implication of our results is that information displays that help more strategically sophisticated agents anticipate others' responses can lead to a differential advantage, resulting in less total welfare across the population but simultaneously reducing distribution shift. While providing an interface that communicates realized prediction errors to the entire \textit{user population} appears to be the optimal default to maximize utility at the individual level, our findings suggest that chasing this objective does not necessarily lead to greater overall social welfare. In fact, interfaces that did not disclose realized prediction error were more effective in maximizing the total potential social welfare of the system. However, deploying an interface visualizing realized prediction errors, despite its potential negative impact on the system's social welfare, can reduce distribution shift. This reduction decreases the disparity between the predicted and realized outcomes, thereby enhancing the perceived accuracy of the information display for agents after decision-making. Consequently, it fosters improved trust and reliance on the information display in the context of repeated strategic decision-making.

Our findings suggest that there are not only ethical considerations that arise in negotiating the trade-off between individual-level utility and aggregate-level social welfare but also an important trade-off between social welfare and users' potential trust and reliance on the information display.

\subsection{Challenges of Estimating Display Effects in Strategic Settings}
We study a scenario where predictions from an information display are based on historical data, which may not accurately reflect the behavior that ensues when agents interact with the display. Rather than trying to reduce the disparity between the predicted and realized outcomes, our interest lies in understanding how different design features of an information display impact individual-level decision-making, aggregate-level system outcomes, and their stability over repeated decisions over time. Given the intricate relationship between these two types of measurements, studies like ours take on a challenging problem.

We believe that our goals serve as an important complement to approaches that aim to ``prescribe" desirable aggregate-level system outcomes through displays. Such approaches include information design, which involves selective information disclosure (e.g.,~\cite{das2017reducing}); the concept of visualization equilibrium~\cite{kayongo2021visualization}  where the displayed prediction matches the realized outcome, essentially represented by a fixed point of the agents' behavior and the visualization function that produces the visualization; and theoretical solutions in machine learning research for addressing distribution shifts induced by predictive algorithms in social prediction contexts (i.e., performative prediction), which involves defining stable or optimal fixed points of retraining against expected future outcomes (e.g.,~\cite{perdomo2020performative,mendler2020stochastic,miller2021outside}). Attempts to inform design with knowledge of causal effects of display characteristics, as we pursue, tend to require fewer assumptions. However, this does not mean that identifying such effects experimentally is easy. The small number of trials we conducted imposes limitations on the dynamics we can observe. Future work might explore behavior over a longer duration to understand patterns of performance improvement and convergence across interfaces with extended usage.

\subsection{Limitations and Future Work}
Our proposed staged experimental design could be applied to a wide range of non-cooperative games under the level-$k$ framework, including those with different payoff structures, asymmetric information, and dynamic environments. However, this design is constrained by the level-$k$ framework, as it evaluates agents' decisions based on a level-specific outcome that aligns with their endowed belief (see Section \ref{sec:level-specific-feedback}). Our design reinforces level endowment by making players myopic, which leads to the consequence that, with this setup, we are unable to capitalize on players' intrinsic strategic sophistication, nor does it explore the dynamics of decision-making and system outcomes when players' levels of sophistication evolve over repeated games in light of the results of previous plays. Future research could aim to infer the level distribution of users in real-world prediction interfaces like driving direction apps or advertising dashboards, so as to complement algorithmic solutions to problems like distribution shift with greater knowledge of how outcomes can arise. Future work might also pursue the use of synchronous experiments using platforms such as empirica.ly~\cite{almaatouq2021empirica} where it is possible for post-decision feedback to combine decisions from all participants in real-time (e.g., \cite{frey2013cyclic}) to avoid the limitations of our staged design.  

Our experiment also did not directly measure reliance on the information display, but an interesting avenue for future work would be to design a similar experiment where an agent's continued interest in using an information display is an outcome variable. For example, an experiment might elicit and quantify their willingness to pay for a display relative to the utility-optimal amount. 

Finally, we used node-link diagrams to present predictions and employed NetHOPs for salient uncertainty quantification. This design choice was partially motivated by the structure of our congestion game, which comprises three districts forming a traffic network that can be most naturally visualized using a graph. However, alternative visualization methods, such as bar charts, might sometimes be preferred due to their simplicity or familiarity. Future research could investigate the sensitivity of the results to the use of different visualization techniques.
\section{Conclusions}
The distribution shift that occurs when presenting predictions in strategic settings poses distinct design challenges for interfaces supporting data-driven decision-making. Through several extensive online experiments in a repeated congestion game, we investigate the impact of design choices, such as visualizing prediction uncertainty and error on both individual decision-making and aggregate system outcomes. By endowing agents with varying level-$k$ depths of thinking from a Poisson-CH model~\cite{camerer2004cognitive} in our innovative staged experimental design, we demonstrate the differential effects of information display based on agents' strategic sophistication. We find that interfaces providing post hoc decision feedback, which visualize realized prediction error, aid more sophisticated agents (L2s) in best responding and anticipating other agents' decisions. Our study also highlights the inherent trade-offs that can arise from enhancing individual decision-making or the overall collective outcome. We observe that while design choices favoring individual-level decision-making may lead to suboptimal collective outcomes in terms of social welfare, they simultaneously enhance user trust and reliance on the prediction interface by reducing distribution shift and hence improve perceived accuracy post hoc. This observation opens up new avenues for research into robust design strategies in scenarios where shared information displays might retrospectively be inaccurate. Our work sets the stage for further exploration into effectively balancing individual utility and collective welfare in strategic environments, thereby influencing future approaches to interface design in shared information settings.

\begin{acks}
This work was supported by NSF IIS-2211939.
\end{acks}

\bibliographystyle{ACM-Reference-Format}
\bibliography{reference}

\appendix
\section{Full Description of Data Pre-processing and Counterfactual Model}\label{sec:strategic-setting-appendix}

We construct our congestion game using the Chicago Taxi Trips data\footnote{The Chicago Taxi Trips data used in this study is publicly accessible via \href{https://data.cityofchicago.org/Transportation/Taxi-Trips/wrvz-psew}{the Chicago Data Portal}.}, which takes the form of origin-destination (OD) flows. Each observation represents a completed trip, including variables such as taxi ID, start/end timestamps, and pickup/drop-off Community Areas\footnote{Community Areas are delineated with fixed boundaries that remain constant over time and are used for census collection. For a more detailed description, see the information provided by \href{https://www.chicago.gov/city/en/depts/dgs/supp_info/citywide_maps.html}{Chicago's Department of Asset, Information and Services}.} (CAs). Leveraging findings from previous work on modeling taxi trips (e.g., ~\cite{chang2010context, moreira2012predictive, liao2018large, zhang2016framework, davis2016multi}) and concepts from spatial econometrics (e.g., ~\cite{legendre2012numerical, anselin1988spatial, nelson2008detecting, goodchild1992geographical}), we process the taxi trips data and formulate a counterfactual model that can support our strategic setting by computing individual player payoffs as well as the system outcome. 

\subsection{Defining the Action Set Available to Players} \label{sec:action-sets-appendix}
We first define the \textit{action set} for our congestion game, which represents the set of choices or strategies available to each player, by analyzing the average daily pickups by CAs between January 2014 and December 2015. This time range was chosen because it represents a peak period in Chicago taxi demand. Like many other major US cities, the taxi industry in Chicago is experiencing an annual decline due to increased market competition from ride-sharing companies~\cite{jiang2018ridesharing, correa2017exploring}. This peak demand period provides a robust data slice, reflecting a highly active environment, essential for accurately simulating the strategic decisions in our congestion game.

\begin{figure}[ht]
  \centering
  \includegraphics[width=0.45\linewidth]{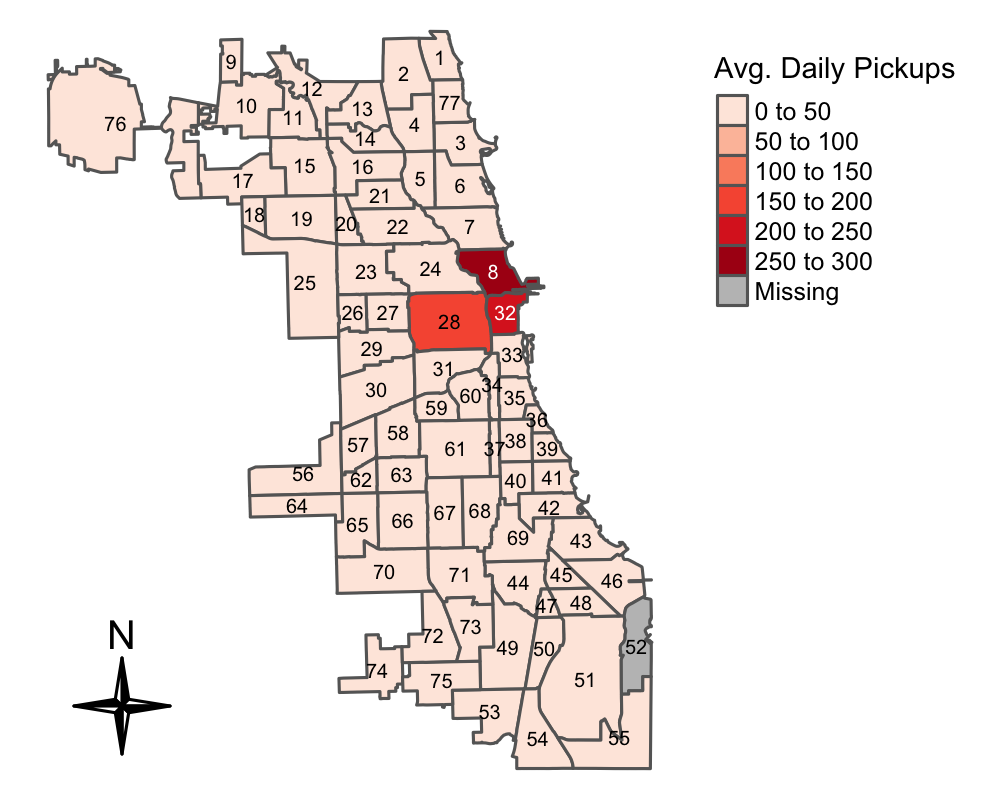}
  \caption{Average daily pickups by Chicago CA.}
  \label{fig:avg-daily-pickups}
  \Description{The averaged number of daily pickups for each of 77 Chicago CAs}
\end{figure}

As shown in \autoref{fig:avg-daily-pickups}, the average daily pickups demonstrate spatial heterogeneity~\cite{anselin1988lagrange}, with the distribution of pickups being highly uneven across various areas. Notably, three CAs in the Central Business District---North Loop (CA = 8), West Loop (CA = 28), and the Loop (CA = 32)---have significantly higher daily pickups compared to the rest of the CAs. Additionally, these three contiguous spatial hot spots~\cite{nelson2008detecting} show evidence of spatial dependency~\cite{goodchild1992geographical}, as indicated by their similar average daily pickups. To manage spatial heterogeneity and limit the number of total actions available in our strategic setting, we define a player's \textit{action set} as consisting of these three spatial hot spots, which are referred to as \textit{action CAs}. 

\subsection{Reducing Heterogeneity in the Training Data} \label{sec:reduce-heterogeneity-appendix}
Taxi pickups can vary considerably depending on the time of day, day of the week, weather conditions, and special events (e.g., sports games, holidays). While real taxi drivers could adjust their use of an information display in light of such factors, in our experimental setting, training a model without accounting for these irregular demand shocks could lead to highly variant display errors. To address this, we implement a stratification scheme that groups taxi trips by start timestamps along two time dimensions: (1) days of the week (i.e., weekdays and weekends) and (2) time intervals of a day\footnote{We consult the time intervals defined by \href{https://movement.uber.com/}{Uber Movement}, which are AM Peak (7 AM - 10 AM), Midday (10 AM - 4 PM), PM Peak (4 PM - 7 PM), Evening (7 PM - 12 AM), and Early Morning (12 AM - 7 AM).}. 

We define the pickup session of our strategic setting to take place at 9 AM, and select the strata containing 5.8 million trips from 522 weekday AM Peaks (7 AM - 10 AM) as the relevant trips to infer each driver's 9 AM search decisions. We also account for two external factors that can influence taxi flows within the strata: weather conditions and the occurrence of special events. For weather conditions, we rely on four hourly meteorological variables\footnote{Hourly meteorological variables are collected from \href{https://weatherstack.com}{Weatherstack API}.} (i.e., apparent temperature, precipitation, snow depth, and wind speed). For special events, we consulted a list of Chicago CBD events\footnote{Data is publicly available from  \href{https://www.chicago.gov/city/en/depts/dca/supp_info/events.html}{Chicago's Department of Cultural Affairs and Special Events}.} as well as national holidays. After conditioning on the time period and external factors, we identified 221 (42\%) homogeneous target weekdays from the trip strata.

\subsection{Deducing Search Flows} \label{sec:prior-inference-appendix}
Our objective is to develop a counterfactual model with a functional form $\text{pickups} = f(\text{flow})$, so that it can predict the total number of pickups (i.e., the dependent variable) in each \textit{action CA} at 9 AM, given a discrete flow distribution of drivers going to search over the three CAs (i.e., the independent variable). Although we can aggregate the dependent variable by directly counting the total number of 9 AM pickups for each \textit{action CA} from the taxi data, we face a hard data limitation that threatens the viability of our counterfactual model: the data only provide partial information about successful pickups (i.e., the confirmed trips as recorded in the taxi trips data) or the \textit{total supply}---we do not know the number of drivers that searched an \textit{action CA} and did not get a pickup.\footnote{We also do not observe total demand, but the convention is to use total pickups from a location as a proxy of total demand.} We therefore design an algorithm that can deduce drivers' search decisions using their prior pickup history during AM Peaks.

At a high level, the algorithm first identifies candidate drivers who might choose to search in each \textit{action CA} at 9 AM on a target weekday by considering a pivotal question: ``What does it mean for a taxi driver \textit{to be able to make a} pickup in an \textit{action CA} at 9 AM on that weekday?". To answer this question, the algorithm (1) back-traces the previous drop-off CAs of drivers who found 9 AM pickup in an \textit{action CA}, and (2) computes the amount of idle time between their previous drop-off and the 9 AM pickup, allowing us to approximate the maximum amount of time drivers who obtained a pickup spent searching from their previous drop-off CAs. The output of this procedure is a collection of weighted edges, which we call \textbf{\textit{trace dyads}}, because they track drivers' movements on the target weekday in the form of $D\xrightarrow{T}P$ where $D$ is the previous drop-off CA, $P$ is an \textit{action CA} where a driver found pickup at 9 AM, and $T$ is the weight representing the amount of idle time between previous drop-off and the 9 AM confirmed pickup. For each unique trace dyad by $D$ and $P$, we find the maximum idle time, $t_{max}$, and set $t_{max}$ as the time threshold used to identify candidate drivers who might choose to search in the \textit{action CA} from each unique $D$. For example, suppose the data shows that the maximum idle time for drivers who previously dropped-off in the North Loop before finding a 9 AM pickup in the Loop is 15 minutes on the target weekday. In that case, we include all unique drivers who dropped-off in the North Loop within 15 minutes before 9 AM into our candidate list for that target weekday.

The algorithm then classifies each candidate driver identified by trace dyads into three types according to their 9 AM vacancy status shown by the taxi data. These types are (1) those who successfully found a pickup in an \textit{action CA} at 9 AM, (2) those who successfully found pickups elsewhere at 9 AM, and (3) those who failed to find a pickup and have no trip history in the data at 9 AM. Drivers of type (1) must have searched in one of the \textit{action CAs} and hence are included in the search flow to the \textit{action CA} where they found a pickup. Drivers of type (2) are removed from the candidate list. Drivers of type (3) are of primary interest: we need to deduce where they might have searched and decide if they should be included in the search flow for an \textit{action CA}. 

The algorithm consults each type (3) driver's search prior and tabulates unique consecutive pickups completed by the driver in the AM Peaks of the past ten days prior to the target weekday. From each driver's search prior, we create a collection of \textbf{\textit{search dyads}} expressed as $V\xrightarrow{N}S$ where $V$ is the drop-off CA of the previous trip, $S$ is the pickup CA of the consecutive trip, and the weight $N$ is the number of occurrences of the pickup pattern. A search dyad can be interpreted as follows: in the AM Peaks of the past ten days, given a driver dropped-off in $v$, she found her next pickup in $s$ for $n$ times. Since the collection of search dyads with the same $V$ can be seen as proxies of a driver's conditional search preference from $V$, the algorithm identifies the $S$ with the maximum $N$ as the search CA for the driver.

\subsection{Counterfactual Model}\label{sec:counterfactual-appendix}
The above pre-processing steps result in a training dataset that contains 9 AM flow-pickup pairs on 221 homogeneous weekdays for each candidate CA, summarized from 2.1 million relevant trips completed by $5,109$ Chicago taxi drivers. Each observation in the processed data describes the number of observed pickups in an \textit{action CA}, resulting from the quantity of deduced flow going into the \textit{action CA} on each of the 221 weekdays. We use a Bayesian Multilevel model to estimate pickup distributions and specify the model in Wilkinson-Rogers-Pinheiro-Bates syntax~\cite{pinheiro2017package, wilkinson1973symbolic} as follows:

\begin{algorithmic}[1]
    \State \textit{lpickup} $\sim$ Gaussian($\mu$, $\sigma$) \label{lst:line:likelihood}
    \State $\mu$ = \textit{lflow} + (1+ \textit{lflow} | \textit{CA}) \label{lst:line:linear-model}
\end{algorithmic}

As shown, $lpickup$ is the dependent variable, and $lflow$ is the independent variable; both are log-transformed before model fitting. $CA$ is a categorical variable used as the identifier for \textit{action CAs}. In line~\ref{lst:line:likelihood}, we define the likelihood function of $lpickup$ to follow a Gaussian distribution, parametrized by $\mu$ (mean) and $\sigma$ (scale). In line~\ref{lst:line:linear-model}, we specify a hierarchical model with varying intercepts and slopes. This approach accounts for the fact that, although pre-processing (as described in Appendix \ref{sec:reduce-heterogeneity-appendix}) can reduce the heterogeneity of pickup-flow dynamics within \textit{action CAs}, each \textit{action CA} can still have intrinsically different dynamics converting flows to pickups. The intercept and slopes can co-vary because if an \textit{action CA} has larger average pickups (i.e., a high intercept), it could signal a relatively stronger rate of pickup conversion from flows. Given that the \textit{action CAs} we defined are neighbors and subject to the effect of spatial dependence, the dynamics learned by the model from an \textit{action CA} can be used to improve estimates about the other CAs. Therefore, we want to model the covariance in flow-pickup dynamics as a result of spatial dependence between the \textit{action CAs}, and improve CA-specific estimates through information pooling~\cite{mcelreath2020statistical, sullivan1999introduction}. Detailed model diagnostics and posterior predictive checks can be found in the Supplemental Material. 

\begin{figure}[ht]
  \centering
  \includegraphics[width=0.8\linewidth, keepaspectratio]{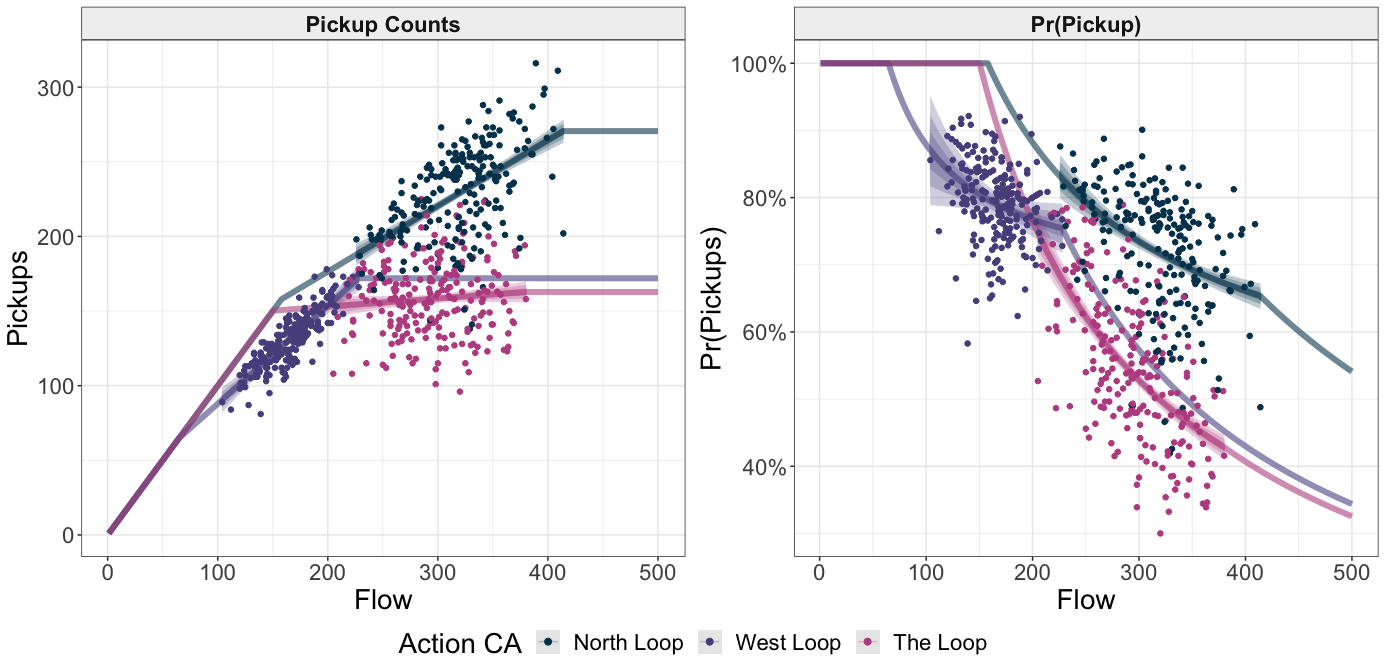}
  \caption{Counterfactual Model Fits}
  \label{fig:modelFits-appendix}
  \Description{Counterfactual Model Fits}
\end{figure}

A challenge in creating a counterfactual model is that the simulated and observed flows passed to the model to generate the display and score decisions may, in some cases, fall below or exceed the flow range observed from the historical data. We use two heuristics to ensure the model makes reasonable predictions in such cases:

\begin{itemize}
    \item If the flow to an \textit{action CA} is greater than the maximum historical flow observed from the data, we use the model to compute an \textit{action CA}'s expected pickups using the historical maximum flows. This approach captures how predicted pickups would eventually stop increasing with greater flow.  
    \item If the flow is less than the minimum historical flow, we use the counterfactual model to predict pickups. Once predicted pickups exceed the simulated flow, we assign a 100\% pickup probability.
\end{itemize}

We present our model fits in \autoref{fig:modelFits-appendix} with the dependent variable in counts (i.e., as defined in the model) on the left and converted pickup probability on the right. When more drivers choose to search in an \textit{action CA}, the pickup probability becomes lower on average, though with varying slopes depending on the CA. The predictions become deterministic only when flows fall outside the range historically observed from the data. 
\section{Decision Scenarios}\label{sec:trial-BR}
\begin{table}[H]
\centering
\caption{
    This table outlines the specific order of weekdays used as decision scenarios in the Main Experiment, the Robust Composition Experiment, and the Robust Trial Order Experiment. It details the best response strategies for each trial, differentiated by participant level and treatment conditions.
    }
\label{tab:trial-orders}
\resizebox{\columnwidth}{!}{%
\begin{tabular}{cccccccc}
\toprule
\multicolumn{2}{c}{Trial Order} & \multirow{2}{*}{Date} & \multirow{2}{*}{Level-1} & \multicolumn{4}{c}{Level-2} \\
Main+Robust Composition & Robust Trial Order &  &  & Static+Bandit & Static+Full & NetHOPs+Bandit & NetHOPs+Full \\
\midrule
0 & 0 & 2014-07-22 & West & East & East & East & East \\
1 & 6 & 2014-05-13 & West & East & East & East & East \\
2 & 14 & 2015-05-21 & West & East & East & East & East \\
3 & 10 & 2014-06-05 & West & East & East & East & East \\
4 & 7 & 2015-06-25 & North & East & East & East & East \\
5 & 12 & 2014-11-07 & West & East & East & East & East \\
6 & 15 & 2015-06-30 & West & East & East & East & East \\
7 & 1 & 2015-10-26 & West & East & East & East & East \\
8 & 9 & 2015-10-22 & West & East & East & East & East \\
9 & 4 & 2015-10-14 & North & East & East & East & East \\
10 & 3 & 2014-09-26 & West & East & East & East & East \\
11 & 8 & 2014-12-05 & West & East, North & East, North & East & East \\
12 & 5 & 2015-08-28 & West & East & East & East & East \\
13 & 13 & 2014-05-22 & West, North & East & East & East & East \\
14 & 2 & 2014-10-24 & West & East & East & East & East \\
15 & 11 & 2014-09-29 & West & East, North & East & East & East\\
\bottomrule
\end{tabular}%
}
\end{table}

\section{Glossary}\label{sec:glossary}

\begin{longtable}{c p{0.65\textwidth}}
\caption{Glossary of terms and concepts used in empirical game theory in the context of our multi-agent strategic setting.}
\label{tab:glossary}\\
\toprule
\textbf{Game Theory} &  \\
\toprule
\endfirsthead
\multicolumn{2}{c}%
{{\bfseries Table \thetable\ continued from previous page}} \\
\endhead
\textit{Congestion game} & A broad class of non-cooperative games where each action represents a congestible good and is associated with a cost function, which incurs cost that increases with the number or fraction of agents who chose the same action. \\
\textit{Principal} & A service or prediction provider. In our experimental setting, we assume the role of the principal, or the taxi company. \\
\textit{Agents or Players} & Users of predictions who also make decisions. In our experimental setting, participants act as taxi drivers who use the information display to inform their search decisions. \\
\textit{Action set} & The collection of all possible actions or strategies available to a player in a game. In our experimental setting, action sets contain three districts, referred to as \textit{action CAs}, that a participant can choose to search for passengers.\\
\textit{Payoff-relevant information} & Information that can directly influence decision-making, thereby determining potential rewards. In our experimental setting, participants' decisions are influenced by two key elements displayed on the information screen: (1) the deduced flow of taxis, and (2) the predicted probability of securing a pickup. \\
\toprule
\textbf{Experimental Setting} &  \\
\toprule
\textit{Poisson Cognitive Hierarchy Model} & A behavioral model we utilize to characterize agents' strategic sophistications through levels, with proportions following a Poisson distribution. \\
\textit{Level-$k$ framework} & In the level-$k$ framework, agents are myopic, believing themselves to be the most sophisticated in action. They perceive other agents as distributed according to a normalized Poisson distribution across levels $0$ to $k-1$. \\
\textit{Level distribution} & A Poisson distribution including L0-L2 drivers representing the ``true'' population mixture over levels, which is exclusive knowledge of the principal used to aggregate the system outcomes by combining decisions using the taxi data and the collected responses from our participants. \\
\textit{Level endowment} & Endowing levels by informing participants of the level mixtures. After normalizing the level distribution, this helps them perceive the rest of the population as consisting entirely of agents of levels lower than their own. \\
\textit{Realized prediction error} & The discrepancy between the predicted outcome and the realized outcome. \\
\textit{Level-specific outcome} & An outcome that aligns with participants' endowed levels to score their decisions and provide feedback in each trial. \\
\textit{System outcome} & An outcome derived by integrating decisions from all levels, in accordance with the level distribution that defines the population. \\
\toprule
\textbf{Experimental Manipulation} &  \\
\toprule
\textit{Static display} & An information display that presents predictions as a static point estimate. \\
\textit{NetHOPs} & A frequency-based uncertainty visualization technique that displays predictions and communicates the associated uncertainty via animated frames. \\
\textit{Bandit feedback} & A feedback mechanism that solely informs participants about whether they secured a pickup based on their decision. \\
\textit{Full feedback} & A comprehensive feedback display that not only informs participants of the decision outcome but also visualizes the realized prediction error. \\
\toprule
\textbf{Response Variable} &  \\
\toprule
\textit{Best response} & A response variable assessing whether a participant chose the district yielding the highest expected pickup probability, as dictated by their endowed level. \\
\textit{Anticipation error} & A response variable evaluating a participant's ability to foresee other participants' search decisions, quantified using the Earth Mover's Distance. \\
\textit{Distribution shift} & A response variable quantifying the discrepancy between deduced and simulated system flows, measured using Earth Mover's Distance. \\
\textit{Social welfare ratio} & A response variable measuring the ratio of total realized pickups to the maximum possible pickups.\\
\bottomrule
\end{longtable}

\end{document}